\documentclass[twocolumn]{aastex62}
\pdfoutput=1 
\usepackage{amsmath,amstext,amssymb,mathtools,graphicx,color}
\usepackage[T1]{fontenc}
\usepackage{apjfonts}
\usepackage[figure,figure*]{hypcap}
\usepackage{chngpage}

\let\splitbox\relax
\usepackage{adjustbox}

\usepackage{graphicx}	
\usepackage{grffile}    
\usepackage{amsmath}	
\usepackage{amssymb}	
\usepackage{xcolor,colortbl}

\definecolor{PineGreen}{HTML}{019286}

\newcommand{\bms}[1]{\boldsymbol{#1}}

\shorttitle{Kilonova Model Grid}
\shortauthors{Wollaeger et al.}
\graphicspath{{./}{figures/}}

\begin{document}

\title{A Broad Grid of 2D Kilonova Emission Models}

\correspondingauthor{Ryan T. Wollaeger}
\email{wollaeger@lanl.gov}

\author[0000-0003-3265-4079]{R.~T. Wollaeger}
\affiliation{Center for Theoretical Astrophysics, Los Alamos National Laboratory,
  Los Alamos, NM, 87545, USA}
\affiliation{Computer, Computational, and Statistical Sciences Division,
  Los Alamos National Laboratory, Los Alamos, NM, 87545, USA}

\author[0000-0003-2624-0056]{C.~L. Fryer}
\affiliation{Center for Theoretical Astrophysics, Los Alamos National Laboratory,
  Los Alamos, NM, 87545, USA}
\affiliation{Computer, Computational, and Statistical Sciences Division,
  Los Alamos National Laboratory, Los Alamos, NM, 87545, USA}
\affiliation{The University of Arizona, Tucson, AZ 85721, USA}
\affiliation{Department of Physics and Astronomy, The University of New Mexico, Albuquerque, NM 87131, USA}
\affiliation{The George Washington University, Washington, DC 20052, USA}

\author[0000-0003-1005-0792]{E.~A. Chase}
\affiliation{Center for Theoretical Astrophysics, Los Alamos National Laboratory,
  Los Alamos, NM, 87545, USA}
\affiliation{Computational Physics Division, Los Alamos National Laboratory, Los Alamos, NM, 87545, USA}
\affiliation{Center for Interdisciplinary Exploration and Research in Astrophysics (CIERA), Northwestern University, Evanston, IL, 60201, USA}
\affiliation{Department of Physics and Astronomy, Northwestern University, Evanston, IL 60208, USA}

\author[0000-0003-1087-2964]{C.~J. Fontes}
\affiliation{Center for Theoretical Astrophysics, Los Alamos National Laboratory, Los Alamos, NM, 87545, USA}
\affiliation{Computational Physics Division, Los Alamos National Laboratory, Los Alamos, NM, 87545, USA}

\author{M. Ristic}
\affiliation{Center for Computational Relativity and Gravitation, Rochester Institute of Technology, Rochester, New York 14623, USA}
\affiliation{Center for Theoretical Astrophysics, Los Alamos National Laboratory, Los Alamos, NM, 87545, USA}

\author[0000-0001-6893-0608]{A.~L. Hungerford}
\affiliation{Center for Theoretical Astrophysics, Los Alamos National Laboratory, Los Alamos, NM, 87545, USA}
\affiliation{Joint Institute for Nuclear Astrophysics - Center for the Evolution of the Elements, USA}
\affiliation{Computational Physics Division, Los Alamos National Laboratory, Los Alamos, NM, 87545, USA}

\author[0000-0003-4156-5342]{O. Korobkin}
\affiliation{Center for Theoretical Astrophysics, Los Alamos National Laboratory, Los Alamos, NM, 87545, USA}
\affiliation{Joint Institute for Nuclear Astrophysics - Center for the Evolution of the Elements, USA}
\affiliation{Computer, Computational, and Statistical Sciences Division, Los Alamos National Laboratory, Los Alamos, NM, 87545, USA}

\author{R. O'Shaughnessy}
\affiliation{Center for Computational Relativity and Gravitation, Rochester Institute of Technology, Rochester, New York 14623, USA}

\author{A.~M. Herring}
\affiliation{Computational Physics Division, Los Alamos National Laboratory, Los Alamos, NM, 87545, USA}

\begin{abstract}

  Depending upon the properties of their compact remnants and the physics included
  in the models, simulations of neutron star mergers can produce a broad range of
  ejecta properties.
  The characteristics of this ejecta, in turn, define the kilonova emission.
  To explore the effect of ejecta properties, we present a grid of 2-component
  2D axisymmetric kilonova simulations that vary mass, velocity, morphology, and
  composition.
  The masses and velocities of each component vary, respectively, from 0.001 to
  0.1 M$_{\odot}$ and 0.05 to 0.3$c$, covering much of the range of results from
  the neutron star merger literature.
  The set of 900 models is constrained to have a toroidal low electron fraction
  ($Y_e$) ejecta with a robust r-process composition and either a spherical or
  lobed high-$Y_e$ ejecta with two possible compositions.
  We simulate these models with the Monte Carlo radiative transfer code SuperNu
  using a full suite of lanthanide and 4th row element opacities.
  We examine the trends of these models with parameter variation, show how it
  can be used with statistical tools, and compare the model light curves and
  spectra to those of AT2017gfo, the electromagnetic counterpart of GW170817.

\end{abstract}

\keywords{Transient sources (1851) --- Infrared sources (793) --- Radiative transfer simulations (1967)
  --- Neutron stars (1108) --- R-process (1324)}

\section{Introduction}

The ejecta from neutron star mergers have long been believed to be a source of
r-process elements~\citep{lattimer74,lattimer76,symbalisty82,eichler89,davies94}.
Whether or not these mergers dominate the r-process elements observed in the
Galaxy depends on the amount of material ejected and the rate of mergers.
The rate of mergers continues to evolve as gravitational wave observations
continue~\citep{abbott20}.
Observations of GW170817 predicted a range of r-process yields depending upon
the analysis of these observations and the nature of the simulations used to
infer ejecta masses from the observations~\citep{cote2018}.
Similarly, estimates of ejecta masses from merger calculations vary considerably
depending upon both the importance of different ejecta mechanisms and the
properties (e.g. masses, spins) of the binary components.
This paper presents the emission from a broad range of ejecta properties to
facilitate more detailed comparisons to observations of these mergers.
The current variation in  ejecta properties depend both upon aspects of the
merger models (theoretical uncertainties) and the initial conditions of the
merging binary (variations expected in nature).
Until the former is constrained better, our range of models must include the
range expected by both.

Simulations of the merger and post-merger environment of binary neutron stars
and neutron star-black hole binaries have suggested a large number of mass
ejection mechanisms, including: tidally disrupted dynamical ejecta, post-merger
MHD and viscosity driven winds from the remnant system (see~\citet{shibata2019}
and references therein), shock-driven ejecta (see, for instance,~\citet{radice2018}),
and cocoon outflow around the GRB jet~\citep{gottlieb2018}.
The different results produced by different calculations depends, in part, upon
how well the simulation technique captures these ejecta processes.
The ejecta mass from merger and post-merger simulations vary over from less than
0.001\,M$_\odot$ to nearly 0.1\,M$_\odot$
\citep{dietrich2017,bovard2017,fujibayashi2018,fahlman2018,radice2018,most2019,shibata2019}.
The masses of the binary objects have a large impact on the amount of mass
ejected~\citep{dietrich2017}.
The mass of the dynamical ejecta and the properties of the remnant also
depend strongly on the neutron star equation of state (EOS)
\citep{sekiguchi2015,dietrich2017} where the remnant lifetime in turn affects the
post-merger wind properties.
Simulations using full general relativity and soft EOS tend to produce lower
tidally-driven dynamical ejecta mass (compare, for instance,
\citet{rosswog2013} and \citet{sekiguchi2015} -  note that these two calculations
use different hydrodynamic methods as well and it is possible that the
hydrodynamic scheme also produces different ejecta masses).

Similarly, the simulated ejecta velocities also vary between different research
groups, with average velocities lying between 0.1 and 0.6 times the speed of
light, $c$ (see GW170817 estimates, for instance~\citet{kasen2017,kawaguchi2018}).
The velocity determines the degree of Doppler broadening of line features in the
spectra, as well as the temperature and density at a given time.
Consequently, differences on the order of $\sim0.05-0.1c$ can have a large
effect on the spectrum at a given time (see, for instance, Figure 3 of~\cite{kasen2017}).

Finally, the composition also varies considerably between different models.
Ejecta with low electron fractions produces a robust heavy r-process, but many
of the ejection mechanisms produce ejecta that consists of a broad range of
electron fractions depending on the nature of the ejection process and the
evolution of the merged object.

All of these properties (mass, velocity and composition) alter the
electromagnetic signature from these mergers.
At this point in time, no perfect simulation exists that captures all of the
physics correctly and the variation between models reflects the assumptions in
the different simulations.
In addition, even if simulations converged on the properties of this ejecta for
a particular progenitor binary, the range of progenitor properties (e.g. spin
and masses of the merging neutron stars) will also produce a wide variety of
ejecta properties.
All of the ejecta properties can be simplified into a 2-component model
(see~\citet{korobkin2020} and references therein) that includes one component
that is very neutron-rich (primarily produced in the tidal ejecta from the
initial merger) and a less neutron-rich component from disk winds, shock driven
ejecta, etc.  We will refer to these two components as low- and high-$Y_e$
respectively.

The differences in ejecta properties seen both in simulations and in the
analysis of GW170817 have motivated a broad set of additional studies of both
ejecta morphology, composition, masses, and velocities
\citep{fontes2020,tanaka2020,kawaguchi2020,kruger2020,korobkin2020,barnes2020,heinzel2020}.
These sets of detailed kilonova models that cover a broad range of the parameter
space are essential in the analysis of kilonova observations.
Here we describe a grid of 900 2D kilonova models, simulated with the Monte
Carlo radiative transfer code {\tt SuperNu}~\citep{wollaeger2013,wollaeger2014}
intended for use by observers to characterize properties of observed kilonovae.
These models vary mass, velocity, high-$Y_e$ morphology, and high-$Y_e$ composition.
The model data used in this work are available from the LANL CTA website.\footnote{\url{https://ccsweb.lanl.gov/astro/transient/transients_astro.html}}
This paper is organized as follows. In Section~\ref{sec:cnms}, we briefly review
the software used to simulate the models.
In Section~\ref{sec:mejp}, we summarize the properties and naming convention for
the model outflows in which the radiative transfer is simulated.
In Section~\ref{sec:rsts}, we verify the anticipated variation of light curves
with model parameter variation, apply some example statistical tools to the model
grid, and compare to the observed light curves and spectra of AT2017gfo, the
electromagnetic counterpart to GW170817.

\section{Codes and Numerical Methods}
\label{sec:cnms}

Our model light curves and spectra are produced with the {\tt SuperNu} radiative
transfer software using tabulated binned opacity~\citep{fontes2020} from the Los
Alamos suite of atomic physics codes~\citep{fontes2015}.
For the composition and radioactive heating from r-process elements, we use the
nucleosynthetic results from the {\tt WinNet} code~\citep{winteler2012}, along
with a decay network to determine the partitioning of energy among the decay products.
The nucleosynthesis uses the Finite-Range Droplet Model (FRDM)
\citep{moller1995}, so the simulation results do not encompass the
variability from uncertainty in nuclear mass models~\citep{barnes2020,zhu2021}.
We employ the decay product thermalization model of~\cite{barnes2016} and use
grey Monte Carlo transport for the gamma-ray energy deposition
\citep{swartz1995,wollaeger2018}.

For the opacities, we employ the binned approach demonstrated by~\cite{fontes2020}.
The opacity tables (including a full suite of lanthanide and 4th row elements)
are over the same density and temperature grid as of~\cite{wollaeger2018}, where
the opacity values are computed assuming local thermodynamic equilibrium (LTE).
This amounts to assuming Saha-Boltzmann statistics apply to computing ionization and
excitation states (the inline {\tt SuperNu} opacity calculation also assumes LTE).

In {\tt SuperNu}, the radiative transfer is restricted to LTE, where the
emissivity of the matter is simply the Planck function multiplied by the
absorption opacity.
This assumption is built into the Implicit Monte Carlo (IMC) time linearization
that ultimately produces the effective scattering terms (instantaneous
absorption and re-emission)~\cite{fleck1971}.
As in previous studies, the outflow is assumed homologous, where radius grows
linearly with time at a velocity proportional to the radius, following the
velocity grid prescription of~\cite{kasen2006}.
Discrete Diffusion Monte Carlo (DDMC) with opacity regrouping
\citep{cleveland2014,wollaeger2014}) is used to optimize the radiative transfer.
The DDMC Doppler shift has been given a Monte Carlo interpretation and
incorporated into the opacity regrouping framework of~\cite{wollaeger2014},
which improves accuracy by removing one operator split in the equations
(Wollaeger 2021, in prep).
All {\tt SuperNu} simulations are performed in 2D axisymmetric geometry, as in
\cite{korobkin2020}.

\section{Model Ejecta Properties}
\label{sec:mejp}

Although there are a number of ejecta processes, for our models, we simplify the
ejecta into two components:  a low electron fraction ejecta characteristic of
the tidal ejecta that occurs during the initial merger and a high electron
fraction ejecta expected from the other ejection processes (wind, shock and
cocoon ejecta).
Simulations of these high-$Y_e$ outflow mechanisms all predict roughly the same
broad range of electron fractions:  $Y_e\sim0.2-0.5$ (see, for instance,
\citet{perego2014,shibata2019,miller2019}).
At this time, these different ejecta sources are difficult to distinguish.
We note that three-component models may introduce a broader range of
behavior in angular variation of light curves and spectra, which we have not
explored in the scope of this model grid.
In this section we present the model naming convention and the set of properties
used to generate the 900 models in the data grid.

\subsection{Model Nomenclature}
\label{sec:nome}

We label models in the following way: \\\\
{\tt T\_m<MD>\_v<VD>\_<S or P><N>\_m<MW>\_v<VW>} \\\\
where T, MD, and VD are the shape, mass, and velocity of the low-$Y_e$ component,
respectively, and S(P), N, MW, and VW, are the shape, composition, mass,
and velocity of the high-$Y_e$ component, respectively.
Masses are in units of solar mass (M$_{\odot}$) and velocities are in units of $c$.
Table~\ref{tb1:nome} has a summary of the properties that form the parameter space.

\begin{table*}
  \centering
  \caption{Model properties per component and associated values.}
  \begin{tabular}{cc}
    Property & Values \\
    \hline
    Low-$Y_e$ mass & $\{0.001,0.003,0.01,0.03,0.1\}$ M$_{\odot}$ \\
    High-$Y_e$ mass & $\{0.001,0.003,0.01,0.03,0.1\}$ M$_{\odot}$ \\
    Low-$Y_e$ velocity & $\{0.05,0.15,0.3\}$ $c$ \\
    High-$Y_e$ velocity & $\{0.05,0.15,0.3\}$ $c$ \\
    Low-$Y_e$ morphology & Toroidal (T; Cassini oval family~\citet{korobkin2020}) \\
    High-$Y_e$ morphology & Spherical (S) or ``Peanut'' (P; Cassini oval family~\citet{korobkin2020}) \\
    Low-$Y_e$ composition & Robust r-process (see Table~\ref{tb1:abun}) \\
    High-$Y_e$ composition & ``High-'' or ``mid-'' latitude wind (see options ``1" and ``2" in Table~\ref{tb1:abun}) \\
    \hline
  \end{tabular}
  \label{tb1:nome}
\end{table*}

In the sections that follow, we use the terms "component" and "ejecta" interchangeably.
We refer to the toroidal component as "low-$Y_e$" and the late-time non-toroidal
component as "high-$Y_e$".
We use the term "wind" only in reference to the elemental composition of the high-$Y_e$ component.

\subsection{Mass-Velocity Grid}
\label{sec:mvs}

The grid of masses and velocities used in the models is inferred from the range
of values acquired from the literature on merger, post-merger, and kilonova light curve
simulations.
Numerical simulations of tidal disruption produce low-$Y_e$ ejecta masses
from $\lesssim10^{-3}$ M$_{\odot}$~\citep{dietrich2017,shibata2019} to $\sim0.05$
M$_{\odot}$ for a binary neutron star system with a stiff EOS and high mass ratio
\citep{dietrich2017}.
For a black hole-neutron star binary, the ejecta mass range goes up to $\sim$0.1 M$_{\odot}$
(see, for instance,~\citet{kruger2020}).
The post-merger high-$Y_e$ ejecta can achieve a comparable range of values
(see Table 1 of~\cite{shibata2019}).
To capture this range, our grid of mass values increase roughly by a root-decade
from 0.001 to 0.1 M$_{\odot}$.
We apply these five mass values to both the low- and high-$Y_e$ component.
The values of mass are listed in the first two rows of Table~\ref{tb1:nome}.

Simulations of~\cite{dietrich2017} of different merger scenarios (mass ratio
and EOS) also show an average low-$Y_e$ ejecta velocity range with error bars
covering 0.1 to $0.2c$ for the low-$Y_e$ ejecta velocity, consistent with the
0.15 to $0.25c$ (or $\sim0.3c$ for prompt black hole formation) range of average
velocity reported by~\cite{shibata2019}.
The range of possible values for the post-merger ejecta is comparable; some models of
GW170817 use a faster blue component around 0.2 to $0.3c$~\citep{kasen2017} while
others have $\sim0.06-0.1c$~\citep{kawaguchi2018,miller2019} for the fiducial model.
The set of average velocities we simulate are consequently from 0.05 to 0.3$c$
for each of the two components, shown in the 3rd and 4th rows of Table~\ref{tb1:nome}.

\subsection{Geometry}
\label{sec:geom}

Ejecta morphology has been shown to alter the nature of the kilonova emission
\citep{korobkin2020}.
Ab initio simulations of the merger produce significant tidally disrupted
ejecta that is toroidally focused (see, for instance,~\citet{rosswog2013}),
and here we assume the low-$Y_e$ dynamical ejecta is toroidal (labeled T) in morphology.
Depending on EOS and the initial binary properties, simulations can produce
more spherical dynamical ejecta from shocks after tidal disruption
\citep{radice2018}, but we do not consider spherical low-$Y_e$ ejecta.
The high-$Y_e$ ejecta morphology is less well-constrained, but calculations of
winds suggest the morphology has a lobed (axially focused or ``peanut shaped'',
labeled P) morphology~\citep{miller2019,korobkin2020}.
However, we include a more traditional spherical morphology (labeled S) for this
late-time ejecta as well in our studies.
We note that a range of high-$Y_e$ ejecta velocities have emerged in simulations
\citep{shibata2019}, and it has been argued that the smoothness of the early
blue spectra of AT2017gfo indicates a fast high-$Y_e$ component surrounding a
slower low-$Y_e$ component (see, for instance,~\citet{kasen2017}).
The portion of our grid of models with slow low-$Y_e$ and fast high-$Y_e$
components adheres to this sort of configuration.
The morphologies used for each component are listed in Table~\ref{tb1:nome}.

The T and P morphologies are generated using the same Cassini oval prescription as
in~\cite{korobkin2020}, and the S morphology is generated with the semi-analytic
spherical formulae provided by~\cite{wollaeger2018}.
Variation over the mass-velocity grid, summarized in Section~\ref{sec:mvs}, does not
change the intrinsic morphology of the ejecta.
Increasing mass for a particular component uniformly scales up the density everywhere
without changing the velocity coordinates, while increasing velocity stretches the
morphology without changing the total mass.
Both types of variation in the low-$Y_e$ ejecta can act to obscure the wind when
the components are superimposed: increasing mass increases lanthanide partial densities,
and increasing velocity covers more volume and reduces lanthanide density.

\begin{figure}
  \centering
  \includegraphics[width=0.55\columnwidth]{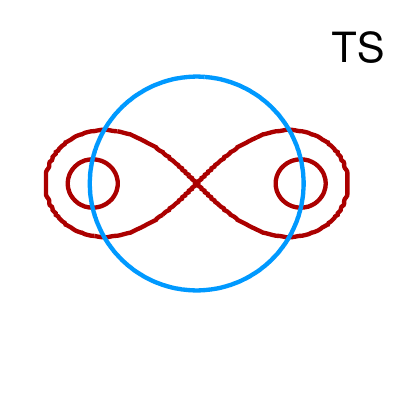}\\
  \includegraphics[width=0.55\columnwidth]{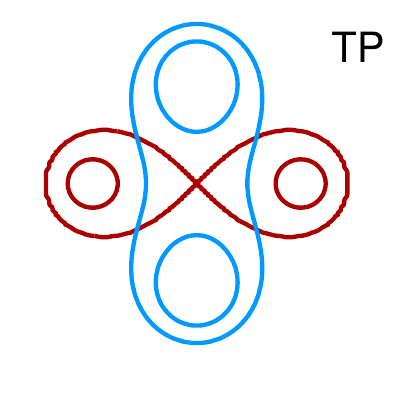}
  \caption{
    Schematics of the two combined morphologies used in the simulation grid
    \citep{korobkin2020}.
    All models have a toroidal (T, red) dynamical ejecta, 450 models are simulated
    a spherical wind (S, blue), and 450 models are simulated with a peanut-shaped
    wind (P, blue).
    Each component is varied over the mass-velocity grid in Table~\ref{tb1:nome},
    and hence is not necessarily drawn to scale here.
  }
  \label{fg1:geom}
\end{figure}

Figure~\ref{fg1:geom} displays schematics of the low-$Y_e$ T morphology (red)
combined with either the S or P-shaped winds (blue), as presented by
\cite{korobkin2020}.
Since each component is varied over the mass-velocity grid in Table
\ref{tb1:nome}, Fig.~\ref{fg1:geom} does not necessarily show the correct
scale of the high-$Y_e$ component relative to the low-$Y_e$ component.
The model morphologies are symmetric under reflection through the (equatorial)
plane bisecting the T morphology, perpendicular to the symmetry axis.
Due to the stochastic nature of Monte Carlo radiative transfer simulations,
equatorial reflection symmetry in the escaping flux is not strictly enforced.

\subsection{Elemental Composition}
\label{sec:abun}

\begin{figure}
  \centering
  \includegraphics[width=0.85\columnwidth]{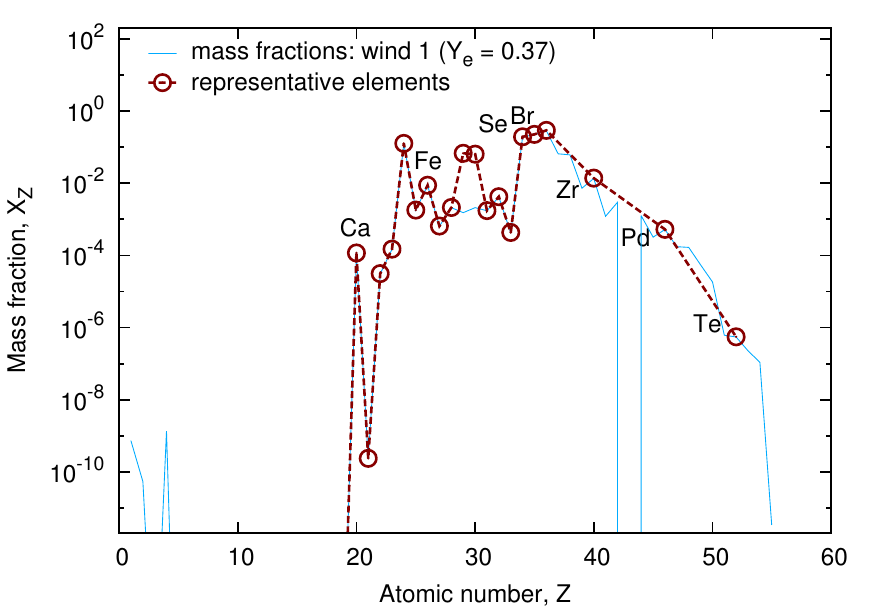}\\
  \includegraphics[width=0.85\columnwidth]{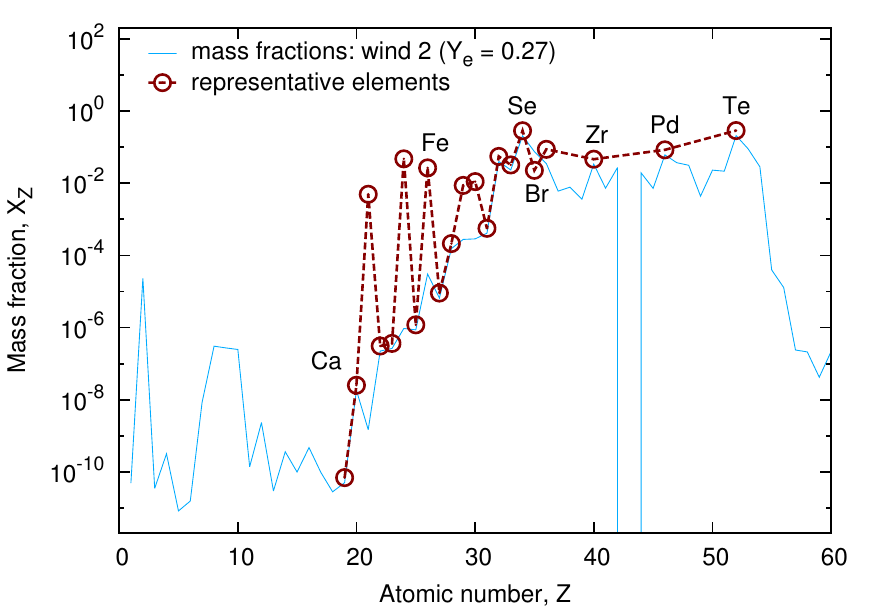}
  \caption{Representative abundances for wind 1 and wind 2 compositions
    (dashed red), including all of the 4th row elements from the periodic table.
    Deviations from original abundances account for elements
    with Z above 4th row: enhanced elements share similar valence
    electron structure with replaced elements.
    These compositions are applied to the high-$Y_e$ components of the
    model grid.
    }
  \label{fg1:abun}
\end{figure}

The remaining ejecta property that significantly affects the emission is
elemental composition.
The compositions of our two components are set by the $Y_e$ and consist of an
r-process composition for the low $Y_e$ ejecta~\citep[as in][]{even2019} and two
different compositions for the high-$Y_e$ ejecta components, implementing a
composition from an averaged $Y_e$ nucleosynthetic yield calculation.
We use high (``wind 1'') and mid (``wind 2'') latitude compositions for the
high-$Y_e$ component, with $Y_e$ values of 0.37 and 0.27 respectively
\citep[see][]{wollaeger2018}.
In \cite{wollaeger2018}, the 4th row elements were represented by a handful of
surrogate elements with calculated opacities.  In these calculations, we include
the full suite of newly calculated 4th row elements with specific contributions
enhanced to account for elements beyond the 4th row that are not included in our
opacity set.
These enhancements were done by comparing valence electron shells between our
4th row element and elements at higher Z (but potentially different principal
quantum numbers).
The elemental abundances are listed for each atomic number in
Table~\ref{tb1:abun}; the wind abundances are plotted versus atomic number in
Fig.~\ref{fg1:abun} (with some of the elements labeled for convenience).

\begin{table}
  \centering
  \caption{Model abundances for the low-$Y_e$ (Dyn. Ej.) and high-$Y_e$
  (Wind 1 or Wind 2) components.}
  \begin{tabular}{|c|c|r|r|r|}
    \hline
    El. & Z & Dyn. Ej. & Wind 1 ($Y_e=0.37$) & Wind 2 ($Y_e=0.27$) \\
    \hline
    K  & 19 & -           & 3.21268e-15 & 6.97642e-11 \\
    Ca & 20 & -           & 1.17432e-04 & 2.52117e-08 \\
    Sc & 21 & -           & 2.38912e-10 & 4.96657e-03 \\
    Ti & 22 & -           & 3.14917e-05 & 3.09675e-07 \\
    V  & 23 & -           & 1.49800e-04 & 3.64843e-07 \\
    Cr & 24 & -           & 1.26889e-01 & 4.73628e-02 \\
    Mn & 25 & -           & 1.81296e-03 & 1.19408e-06 \\
    Fe & 26 & 5.32034e-06 & 8.80579e-03 & 2.65509e-02 \\
    Co & 27 & -           & 6.53211e-04 & 9.01509e-06 \\
    Ni & 28 & -           & 2.12761e-03 & 2.12444e-04 \\
    Cu & 29 & -           & 6.76567e-02 & 8.68511e-03 \\
    Zn & 30 & -           & 6.38433e-02 & 1.10971e-02 \\
    Ga & 31 & -           & 1.74815e-03 & 5.66150e-04 \\
    Ge & 32 & -           & 4.23957e-03 & 5.50677e-02 \\
    As & 33 & -           & 4.33796e-04 & 3.19672e-02 \\
    Se & 34 & 1.01267e-01 & 1.92845e-01 & 2.84281e-01 \\
    Br & 35 & 2.32170e-06 & 2.24753e-01 & 2.27961e-02 \\
    Kr & 36 & -           & 2.89604e-01 & 8.73940e-02 \\
    Zr & 40 & 3.72218e-01 & 1.37605e-02 & 4.64549e-02 \\
    Pd & 46 & 1.38800e-04 & 5.27557e-04 & 8.39116e-02 \\
    Te & 52 & 3.85045e-01 & 5.55957e-07 & 2.88675e-01 \\
    La & 57 & 5.11321e-04 & -           & -           \\
    Ce & 58 & 8.65786e-04 & -           & -           \\
    Pr & 59 & 8.59354e-05 & -           & -           \\
    Nd & 60 & 1.49709e-03 & -           & -           \\
    Pm & 61 & 5.41955e-04 & -           & -           \\
    Sm & 62 & 2.03032e-03 & -           & -           \\
    Eu & 63 & 1.54922e-03 & -           & -           \\
    Gd & 64 & 5.12530e-03 & -           & -           \\
    Tb & 65 & 3.27066e-03 & -           & -           \\
    Dy & 66 & 1.39657e-02 & -           & -           \\
    Ho & 67 & 3.64017e-03 & -           & -           \\
    Er & 68 & 1.10523e-02 & -           & -           \\
    Tm & 69 & 2.34168e-03 & -           & -           \\
    Yb & 70 & 6.44273e-03 & -           & -           \\
    U  & 92 & 8.83764e-02 & -           & -           \\
    \hline
  \end{tabular}
  \label{tb1:abun}
\end{table}

The low-$Y_e$ robust r-process dynamical ejecta composition is unchanged
from~\cite{even2019} which has a simple non-lanthanide composition
relative to the wind compositions.
The impact of the lighter elements on the dynamical ejecta composition
is small due to the effect of ``lanthanide curtaining"
\citep{barnes2013,kasen2015}.
Conversely, the impact of a very small fraction of lanthanides in
the wind 2 composition ($\sim 10^{-7}$) is of little consequence in
light curves and spectra~\citep{even2019}.

As in previous studies, the low- and high-$Y_e$ ejecta compositions are
uniform in each component, and mixed by mass-fraction weighting in the
regions of space where the components overlap.
Consequently, despite wind 1 corresponding to a ``high latitude''
nucleosynthetic tracer, it is in fact applied to all latitudes of the
high-$Y_e$ component morphology (as is the wind 2 composition).

\section{Numerical Results}
\label{sec:rsts}

The effect of each of the model ejecta properties described in Section~\ref{sec:mejp}
on emission can be explored with statistics, given the size of the model grid.
We present some example uses of our model grid,
including basic statistics and comparison with an observation (AT2017gfo), using
data in the expanded form of Table~\ref{tb1:app}.
Specifically, in Section~\ref{sec:vang} we establish the diversity in angular
variation among the models, which complicates the analysis.
In Section~\ref{sec:tren}, we examine trends in means and standard deviations
with variations in model properties and verify consistency with associated
physics.
In Section~\ref{sec:wmcp}, we split the models by morphology and composition of
the high-$Y_e$ component, and explore whether a reduced set of luminosity values can
statistically distinguish the two model populations.
In Section~\ref{sec:at17}, we explore some rudimentary ways to compare the
data set to AT2017gfo.

For all the models, given the geometry, the axial view is least obscured
by the low-$Y_e$ dynamical ejecta and the edge view is most obscured, though the degree
to which the low-$Y_e$ component is obscured depends on the relative speed of
the high-$Y_e$ component.
This is shown in Section~\ref{sec:vang}.
Since the on-axis and edge-on views represent the two extremes of light
curve variation with respect to viewing angle, we restrict the data considered
in Sections~\ref{sec:tren} and~\ref{sec:wmcp} to these views.

\subsection{Viewing Angle Dependence}
\label{sec:vang}

\begin{figure}
  \centering
  \includegraphics[width=0.95\columnwidth]{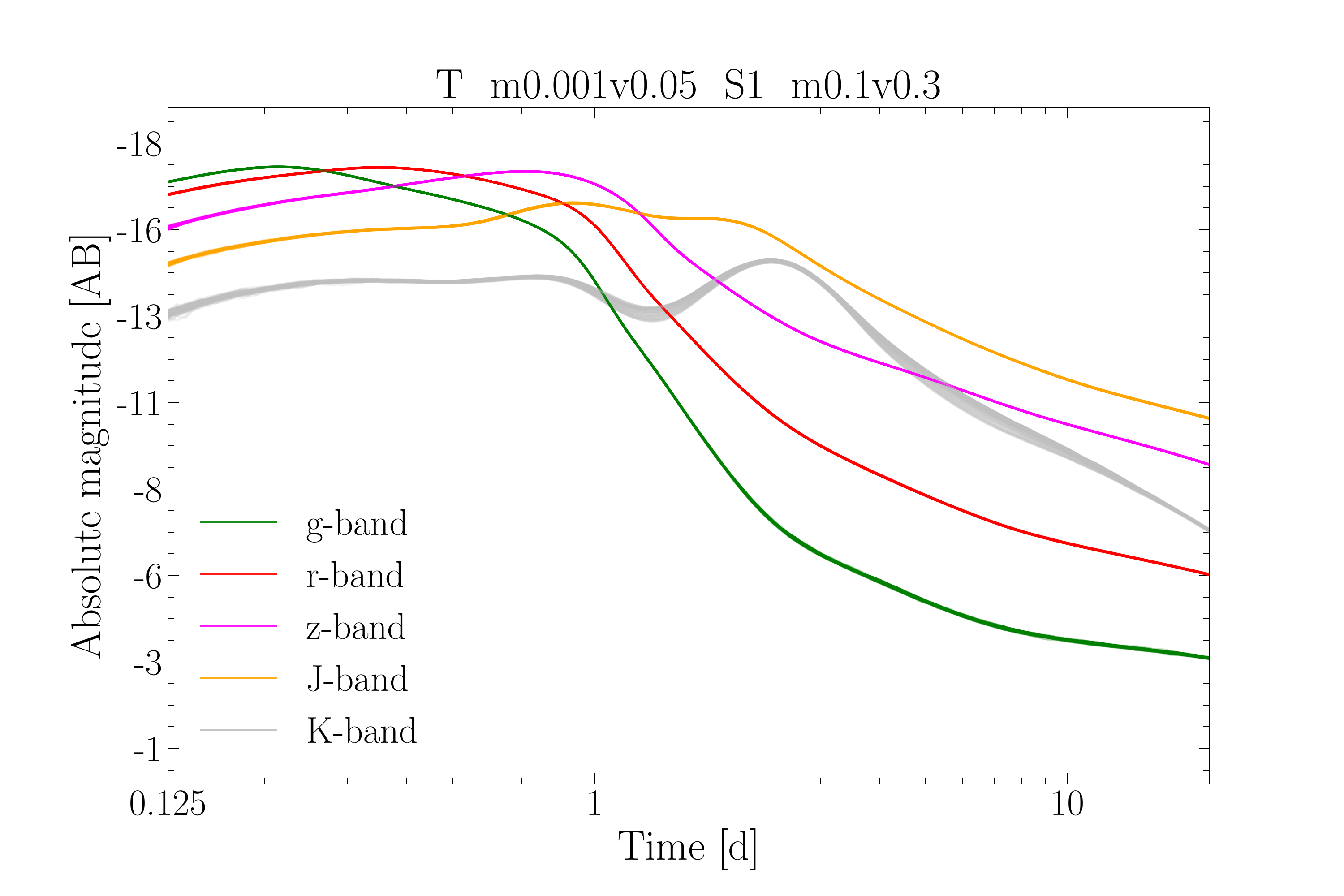}\\
  \includegraphics[width=0.95\columnwidth]{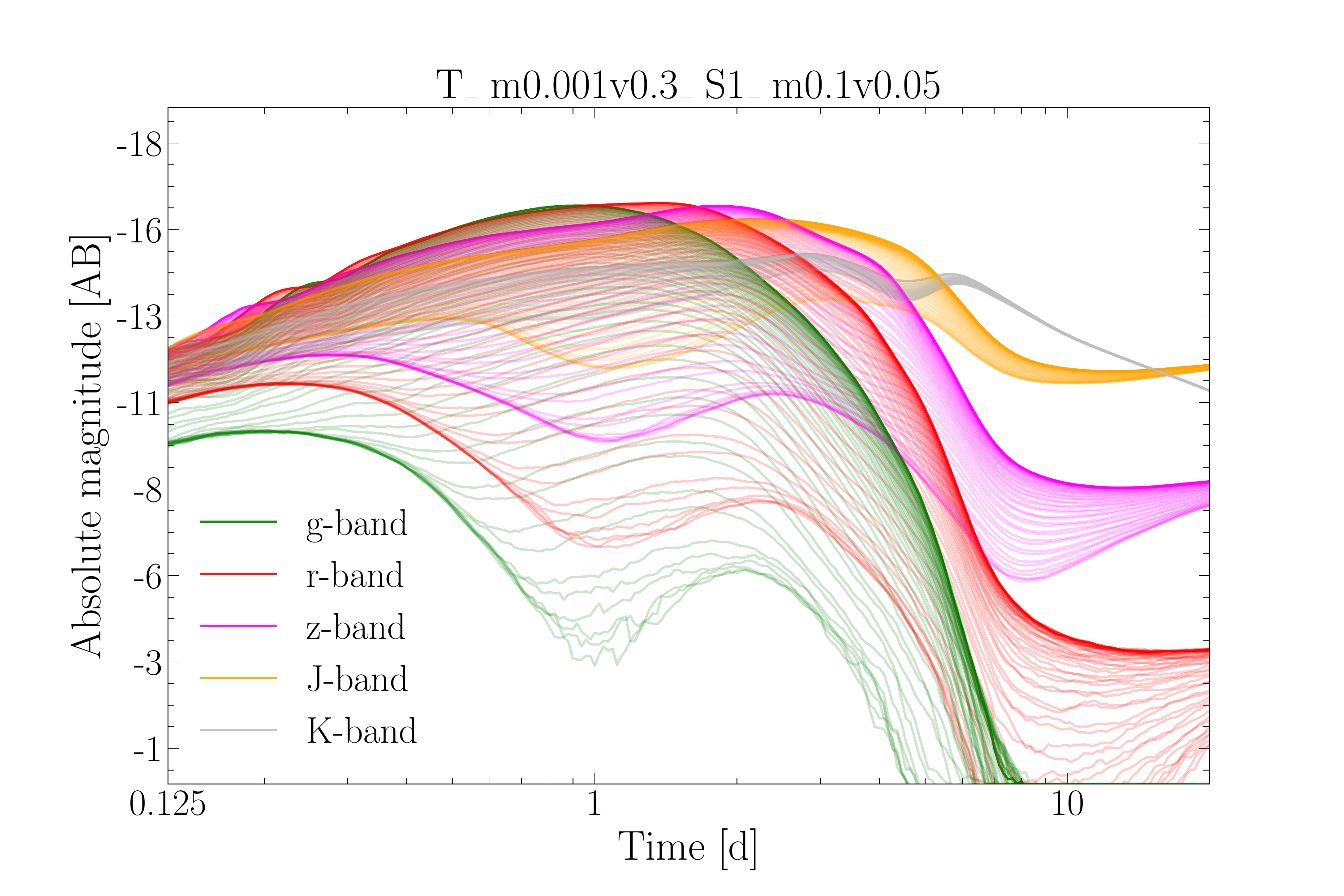}
  \caption{$g$, $r$, $z$, $J$
    and $K$-band absolute AB magnitudes versus time for all
    viewing angles, for models with $0.001$ M$_{\odot}$ and $0.1$ M$_{\odot}$
    in the low-$Y_e$ and high-$Y_e$ component, respectively.
    Top: low-$Y_e$ and high-$Y_e$ ejecta speed of $0.05c$ and $0.3c$, respectively.
    Bottom: low-$Y_e$ and high-$Y_e$ ejecta speed of $0.3c$ and $0.05c$, respectively.
  }
  \label{fg1:vang}
\end{figure}

The degree to which the low-$Y_e$ ejecta obscures the blue kilonova from the
high-$Y_e$ ejecta depends strongly on the relative speeds of the components.
This dependence can be seen in the weak viewing angle variation in models with faster
high-$Y_e$ ejecta and strong viewing angle variation in models with faster low-$Y_e$ ejecta.
This is demonstrated in Fig.~\ref{fg1:vang}, which displays $g$, $r$, $z$, $J$
and $K$-band absolute AB magnitudes versus time for two models with low-$Y_e$
ejecta velocities at alternate extremes in the model set.
Even for the lowest ratio of low-$Y_e$ to high-$Y_e$ ejecta mass, the model with
fast low-$Y_e$ and slow high-$Y_e$ ejecta still produces substantial angular
variation in the magnitudes (bottom panel).
Alternatively, leaving all other properties unchanged, the fast high-$Y_e$ and
slow low-$Y_e$ ejecta (top panel) substantially reduces the angular variation in
the magnitudes.
This phenomenon is the aforementioned lanthanide curtaining effect~\citep{kasen2015},
arising from the relatively high bound-bound contribution to opacity from the
lanthanide abundances in the low-$Y_e$ component (see, for instance,
\citet{gaigalas2019,fontes2020,tanaka2020}).
The diversity of angular variation in emission due to changes in the relative
velocity of the components complicates the categorization of models
based on other properties.
Despite this complicating factor, in the following sections we attempt to establish
expected trends collectively in the models, and determine if model morphology or
composition are discernible as statistically distinct from the emission.

\subsection{Collective Data Trends}
\label{sec:tren}

Various kilonova and supernova light curve studies have established
basic trends in brightness that we expect to be reproduced in our grid of models.
Specifically, we expect that increasing ejecta mass increases brightness and
broadens the light curves in time (at least the bolometric luminosity), increasing
the ejecta velocity increases brightness and narrows the light curves in time (shifting
the peak luminosity to earlier time), and increasing the opacity decreases the brightness
and broadens the light curves in time.
These relationships have been encapsulated in power law relations of luminosity
with respect to mass, velocity, and opacity (see, for instance,
\citet{arnett1979,li1998,grossman2014,wollaeger2018}).

\begin{figure*}
  \centering
  \includegraphics[width=0.9\columnwidth]{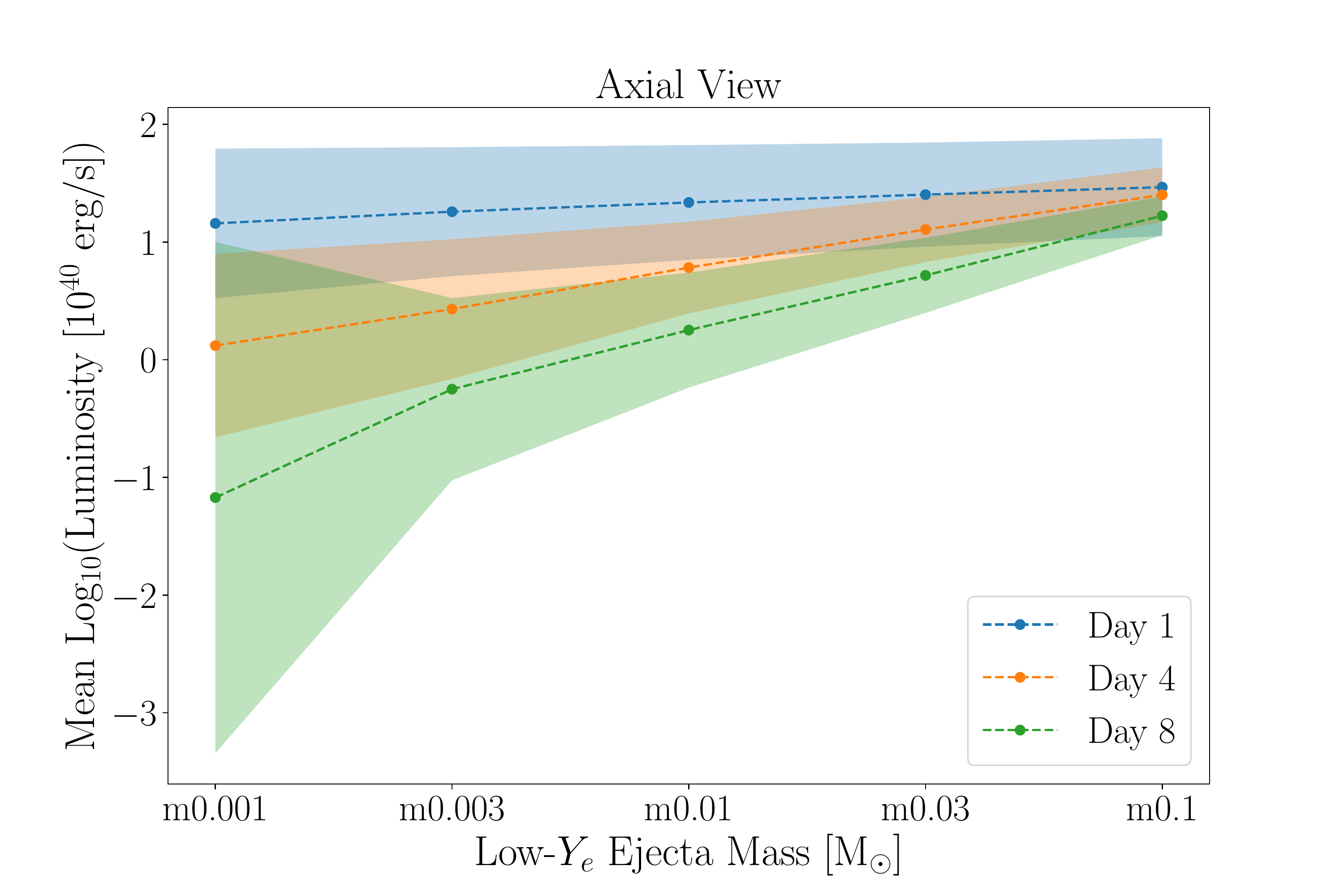}
  \includegraphics[width=0.9\columnwidth]{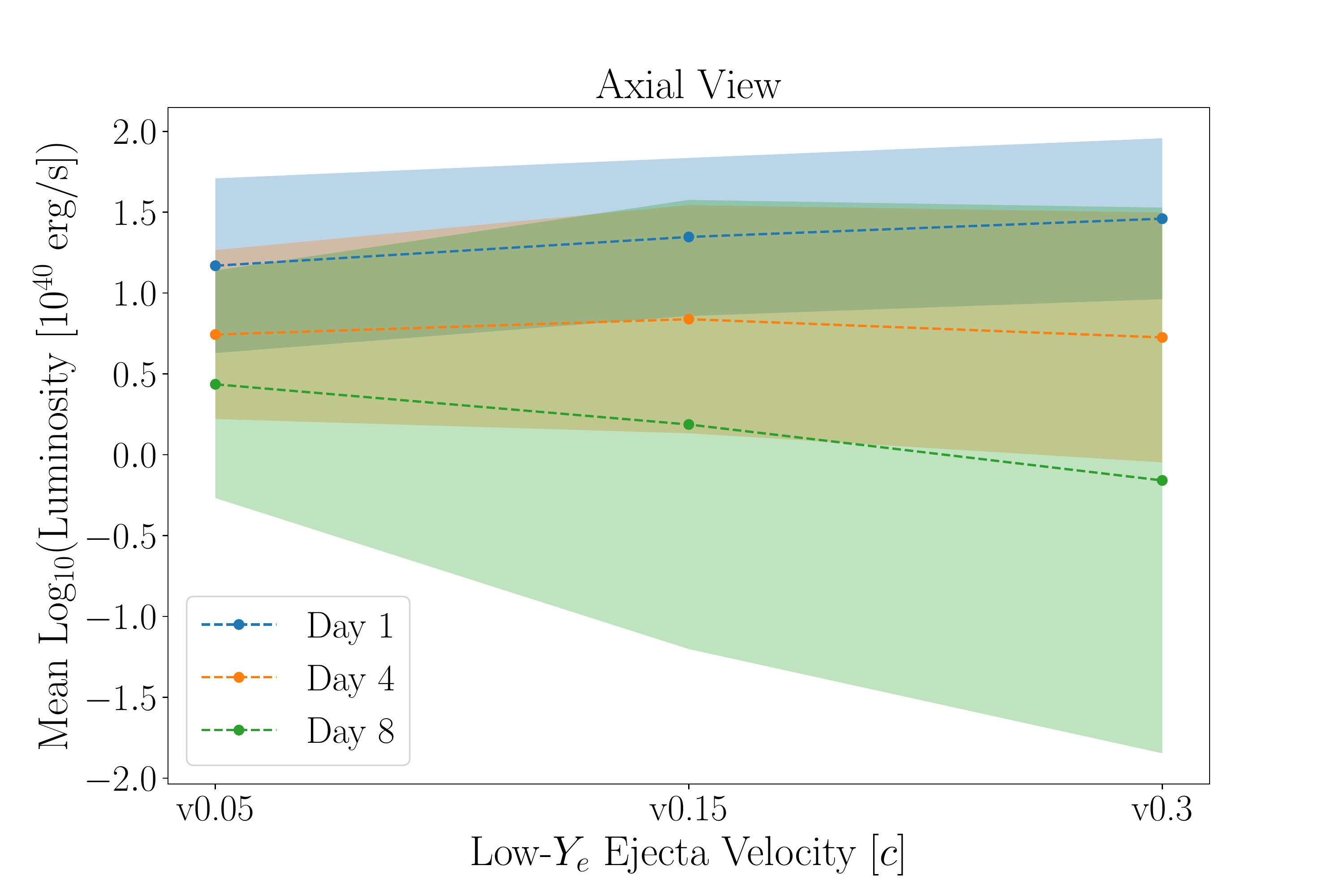} \\
  \includegraphics[width=0.9\columnwidth]{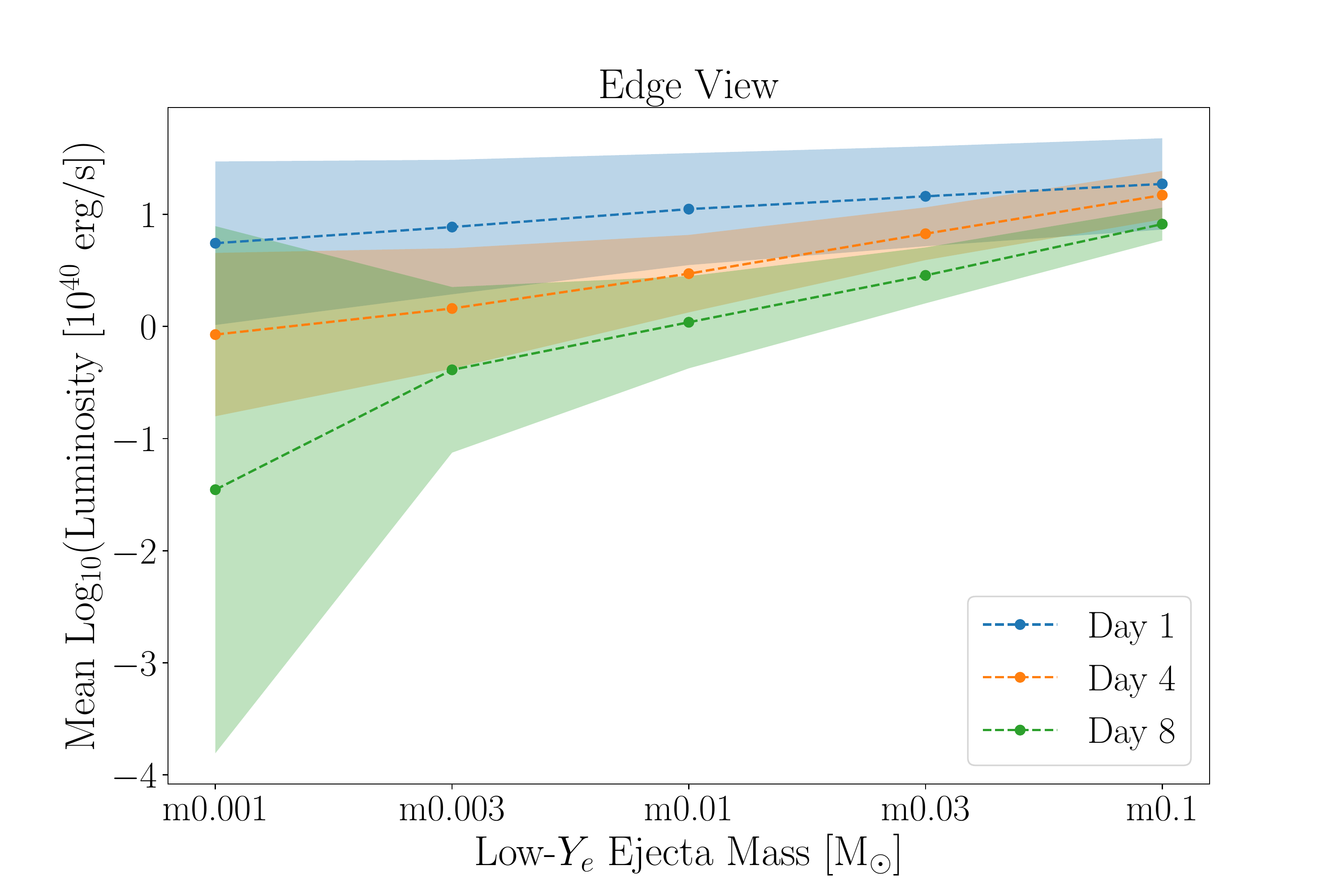}
  \includegraphics[width=0.9\columnwidth]{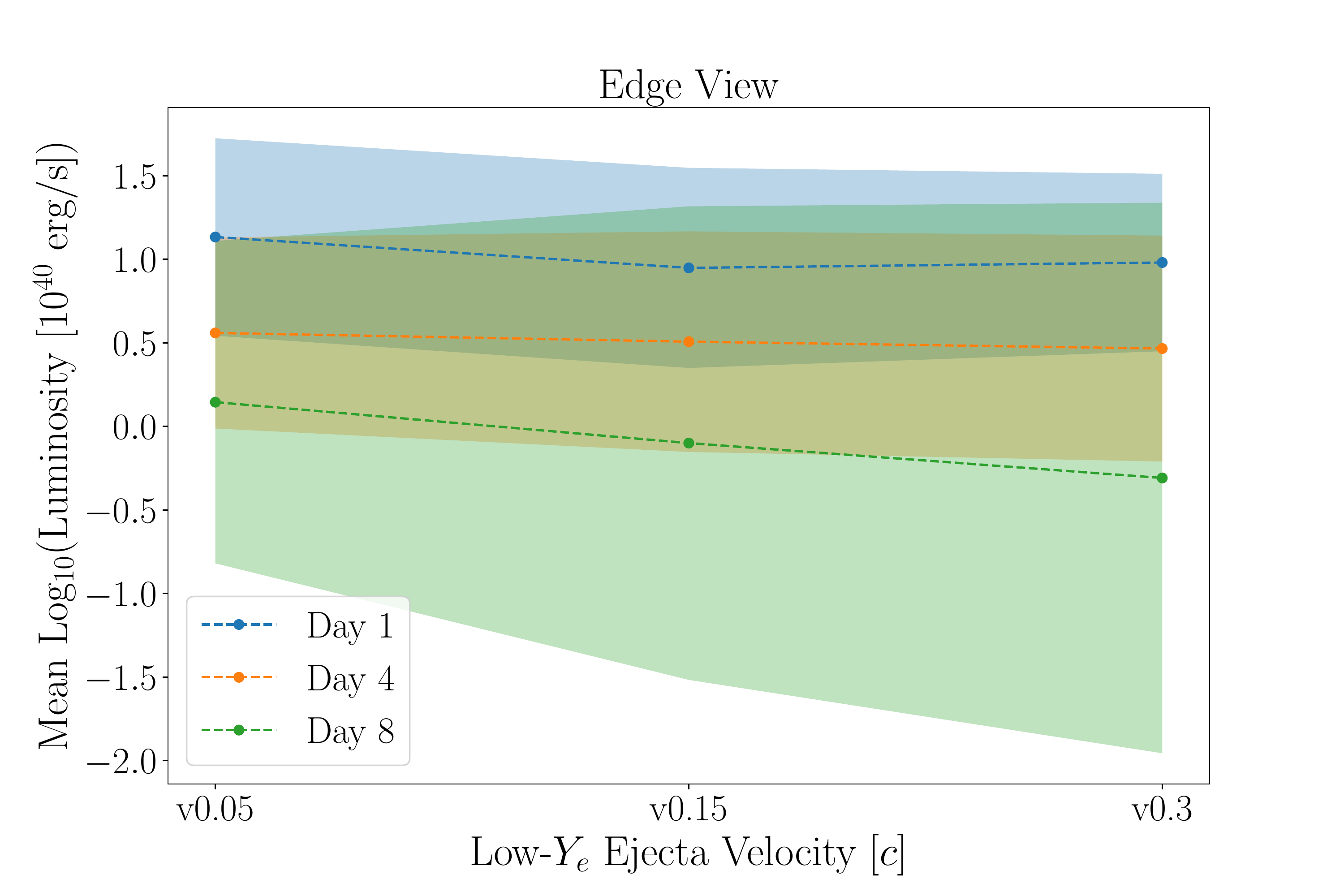}
  \caption{
    Mean of scaled logarithm of bolometric luminosity per low-$Y_e$
    ejecta mass (left panels) and velocity (right panels) for axial (top panels) or
    edge (bottom panels) views at days 1 (blue), 4 (orange), and 8 (green).
    The shaded regions are $\pm$1 standard deviation around each mean
    (colored correspondingly).
    The trends follow the expected patterns: higher mass increases the luminosity
    at each time, and higher velocity tends to decrease the day 1 luminosity in the
    edge view.
    The variance at each value is large at day 1 across properties, but drops at
    day 4 and 8, indicating the low-$Y_e$ ejecta properties become more significant
    in setting the luminosity.
  }
  \label{fg1:tren}
\end{figure*}

Here we verify that our models collectively produce the basic trends expected
in two-component kilonova models with the assumed morphologies.
With a large set of models uniformly spread over the space of parameters,
we may perform some simple statistics to identify trends for certain
parameter variations.
In particular, for the axial and edge observer views, we calculate the
arithmetic mean and standard deviation, per a fixed model property, of the
bolometric luminosity, $g$ and $K$ bands at days one, four and eight.
For luminosity, in Fig.~\ref{fg1:tren} we scale by 10$^{40}$ erg/s and take
the base 10 logarithm of the result in order to improve visibility of the trends,
\begin{equation}
  \mathcal{L} = \log_{10}\left(\frac{L}{10^{40}}\right) \;\;,
  \label{eq1:tren}
\end{equation}
where $\mathcal{L}$ is the scaled log luminosity that is plotted and $L$
is the original luminosity in erg/s.
The day 1, 4, 8 top and side view bolometric luminosity, $g$ and $K$ band
magnitudes for all 900 models are listed in the expaned form of Table
\ref{tb1:app} in Appendix~\ref{sec:app}.
The times are selected to sample the blue kilonova emission from the high-$Y_e$
component (days 1--4) and the red kilonova from the low-$Y_e$ component (days 4--8), and
the $g$ and $K$ bands cover the relevant extremes in wavelength of the spectra.

Figure~\ref{fg1:tren} has day one, four, and eight mean log luminosities versus low-$Y_e$
ejecta mass (left column) and low-$Y_e$ ejecta velocity (right column), with
$\pm$1 standard deviation regions shaded about the means.
The top row of panels is the axial view and the bottom row is the edge view.
The mean and standard deviation are taken for the models over all other properties.
Consequently, each mean or standard deviation at a given mass value is computed
from 900/5=180 points of simulation data at that mass.
From the left column of Fig.~\ref{fg1:tren}, we observe the expected trend of
increasing luminosity with increasing ejecta mass.
Moreover, at later times, the change in increase in luminosity per increase in
mass is more pronounced, indicating that the low-$Y_e$ ejecta mass more strongly
sets the later time luminosity.
Comparing the data at day 1 for the top left and bottom left panels of
Fig.~\ref{fg1:tren}, the luminosity at day 1 is more sensitive to the low-$Y_e$
ejecta mass for the edge view (bottom) than for the axial view, consistent with
the low-$Y_e$ component having a larger effect on edge views on average.
This can be seen by comparing the ratio of the mean luminosity between adjacent
mass values: for the axial view the average increase in brightness is 19\% and
for the edge view the average increase is 36\%.

As expected, the standard deviation of day 1 bolometric luminosity is not as sensitive
to low-$Y_e$ ejecta mass as at day 4 or day 8, for either the axial or edge views.
This result is consistent with the expectation that the early blue transient is not
produced by the low-$Y_e$, lanthanide-rich component, for the parameter ranges studied.
However, it is worth noting that the standard deviation of luminosities at days 1,
4, and 8 are all $\gtrsim$50\% of the total luminosity at the highest mass,
meaning the other ejecta properties still significantly influence the total luminosity
at late time (this notion is consistent, for instance, with radiative emission from the
high-$Y_e$ component being reprocessed into the IR by the dynamical ejecta; the
high-$Y_e$ mass has an effect through this reprocessing).

It is more difficult to ascertain a trend in luminosity with low-$Y_e$ ejecta velocity,
partly due to there being fewer points in the grid of models for velocity.
There appears to be a weak trend at day 1, where the mean luminosity increases
in the axial view as low-$Y_e$ ejecta velocity is increased, and decreases in the edge
view as low-$Y_e$ ejecta velocity is increased.
A physical explanation is that the higher velocity causes higher blue shift of
emission from the dynamical ejecta in the observer frame, supporting a larger
contribution to the total luminosity at early time (before the recession of photosphere).
In contrast, for edge views, the faster low-$Y_e$ ejecta velocity more effectively
obscure any given high-$Y_e$ ejecta, which acts to lower the total luminosity.

\begin{figure*}
  \centering
  \includegraphics[width=0.9\columnwidth]{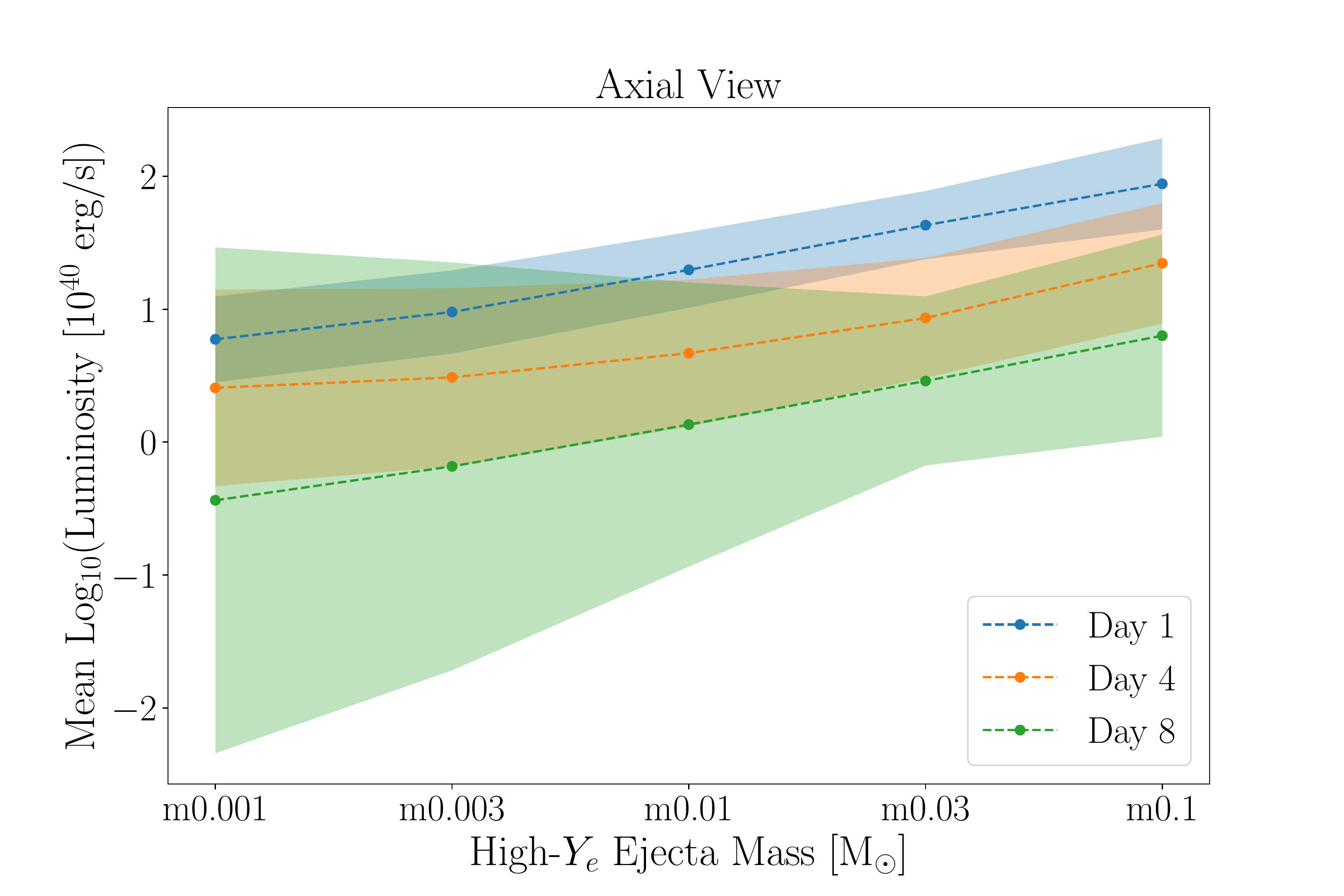}
  \includegraphics[width=0.9\columnwidth]{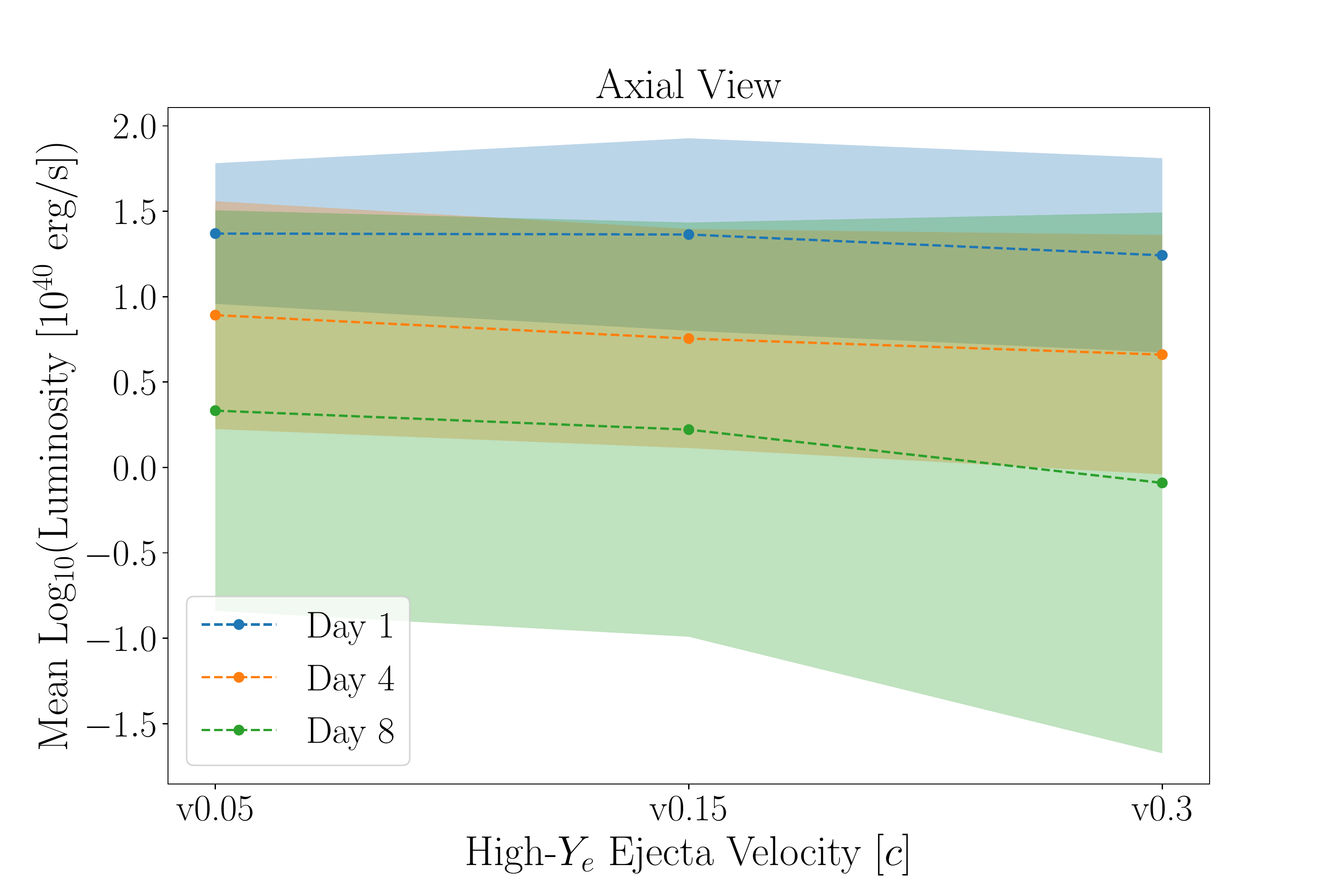} \\
  \includegraphics[width=0.9\columnwidth]{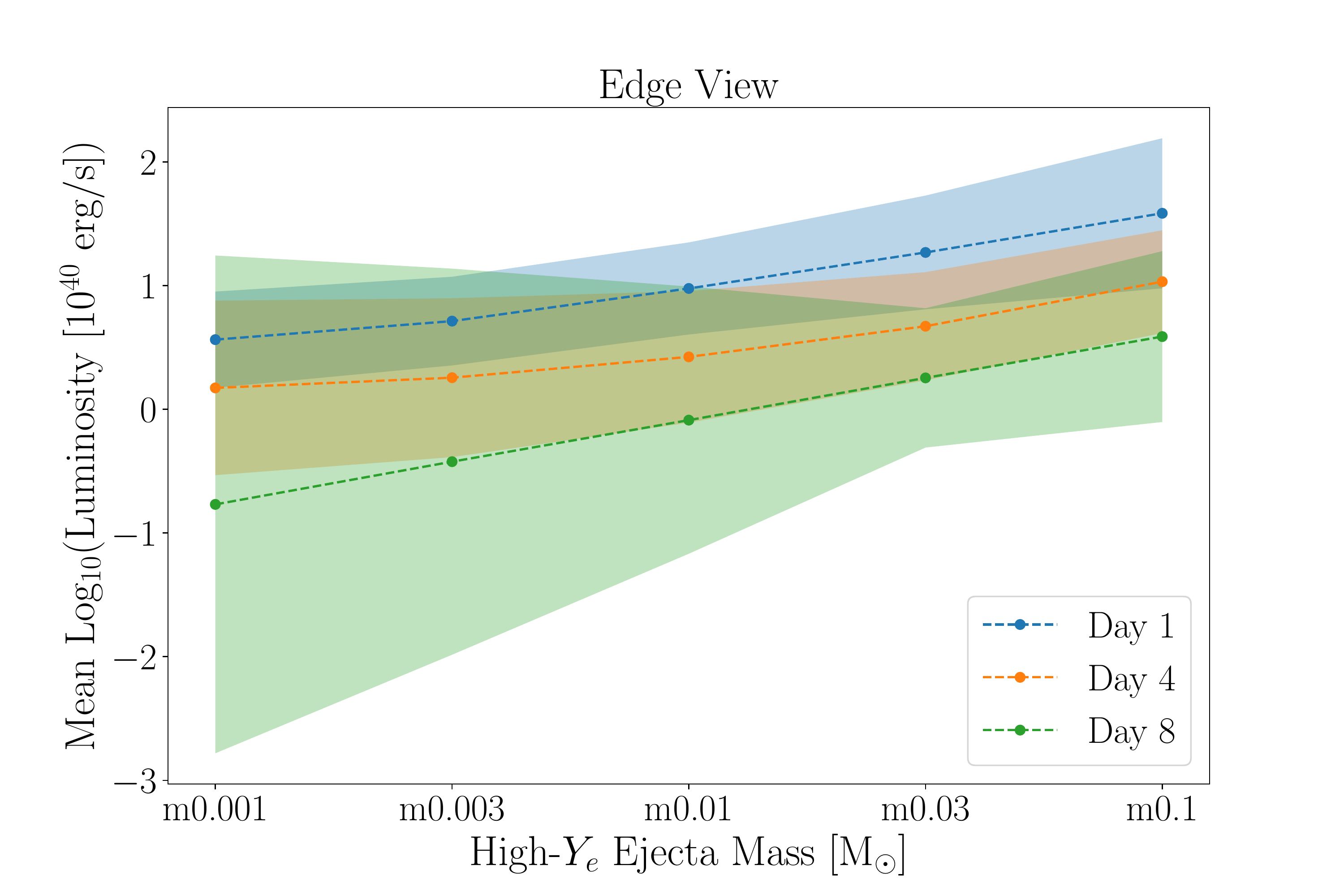}
  \includegraphics[width=0.9\columnwidth]{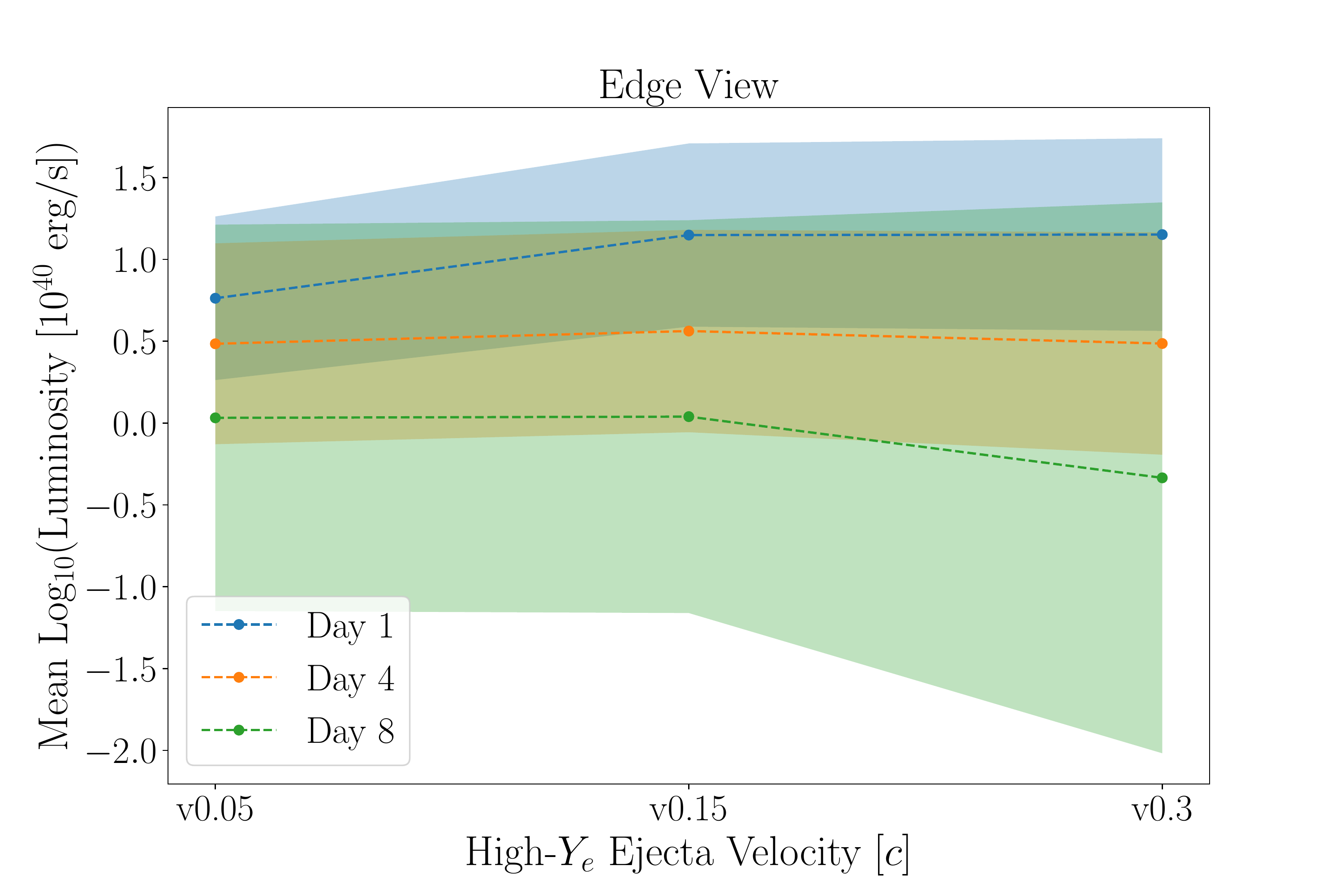}
  \caption{Mean of scaled logarithm of bolometric luminosity per
    high-$Y_e$ mass (left panels) and velocity
    (right panels) for axial (top panels) or edge (bottom panels) views
    at days 1 (blue), 4 (orange), and 8 (green).
    The shaded regions are $\pm$1 standard deviation around each mean
    (colored correspondingly).
    Varying high-$Y_e$ ejecta mass induces a significant trend in the mean
    luminosity at day 1; the trends for later time are consistent with
    reprocessing of high-$Y_e$ emission by the low-$Y_e$ ejecta.
  }
  \label{fg2:tren}
\end{figure*}

Figure~\ref{fg2:tren} has day 1, 4, and 8 mean log luminosities versus
high-$Y_e$ ejecta mass (left column) and high-$Y_e$ ejecta velocity (right column),
with $\pm$1 standard deviation regions shaded about the means.
Relative to Fig.~\ref{fg1:tren}, the bolometric luminosity in both the axial and
edge views is more sensitive to high-$Y_e$ mass, varying by an order of magnitude from
0.001 to 0.1 M$_{\odot}$.
In contrast to the variation with low-$Y_e$ mass, the variation in bolometric
luminosity with high-$Y_e$ mass is greater in the axial view than in the edge
view; for the axial view the average increase in brightness between adjacent
high-$Y_e$ masses is 97\%, and for the edge view it is 82\%.
The relative standard deviation (standard deviation over luminosity) is lowest at
day 1 in the axial view for all masses, which together with the sensitivity of the mean
indicates that the trend is sensitive to fewer other properties than at days 4 and 8.
The high-$Y_e$ ejecta velocity does not appear to have a significant cumulative effect
on the overall brightness, where the most significant change occurs in the edge view
when going from $0.05c$ to $0.15c$; the increase of this change may again be related to
the high-$Y_e$ component becoming less obscured by the low-$Y_e$ component.

\begin{figure*}
  \centering
  \includegraphics[width=0.9\columnwidth]{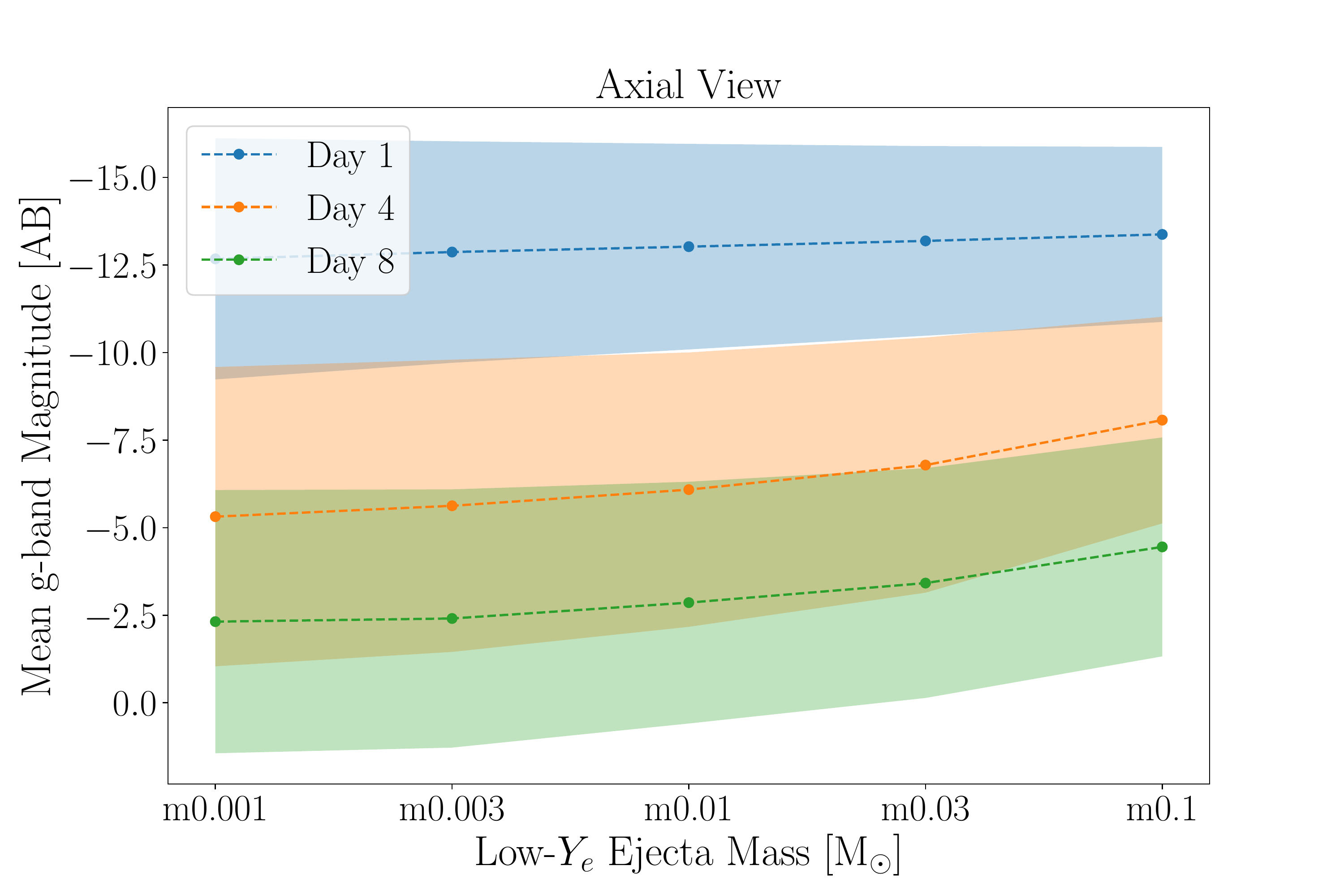}
  \includegraphics[width=0.9\columnwidth]{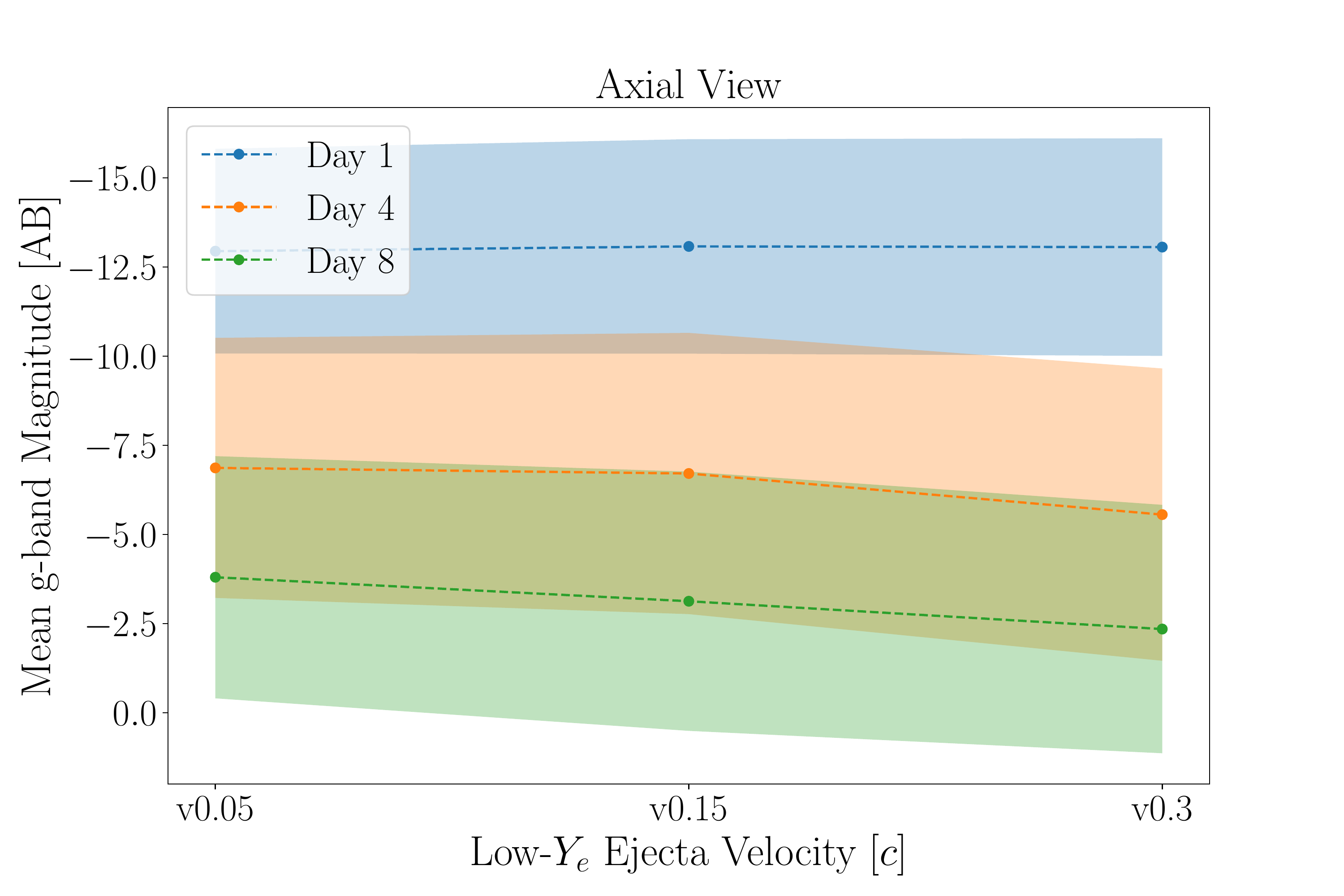} \\
  \includegraphics[width=0.9\columnwidth]{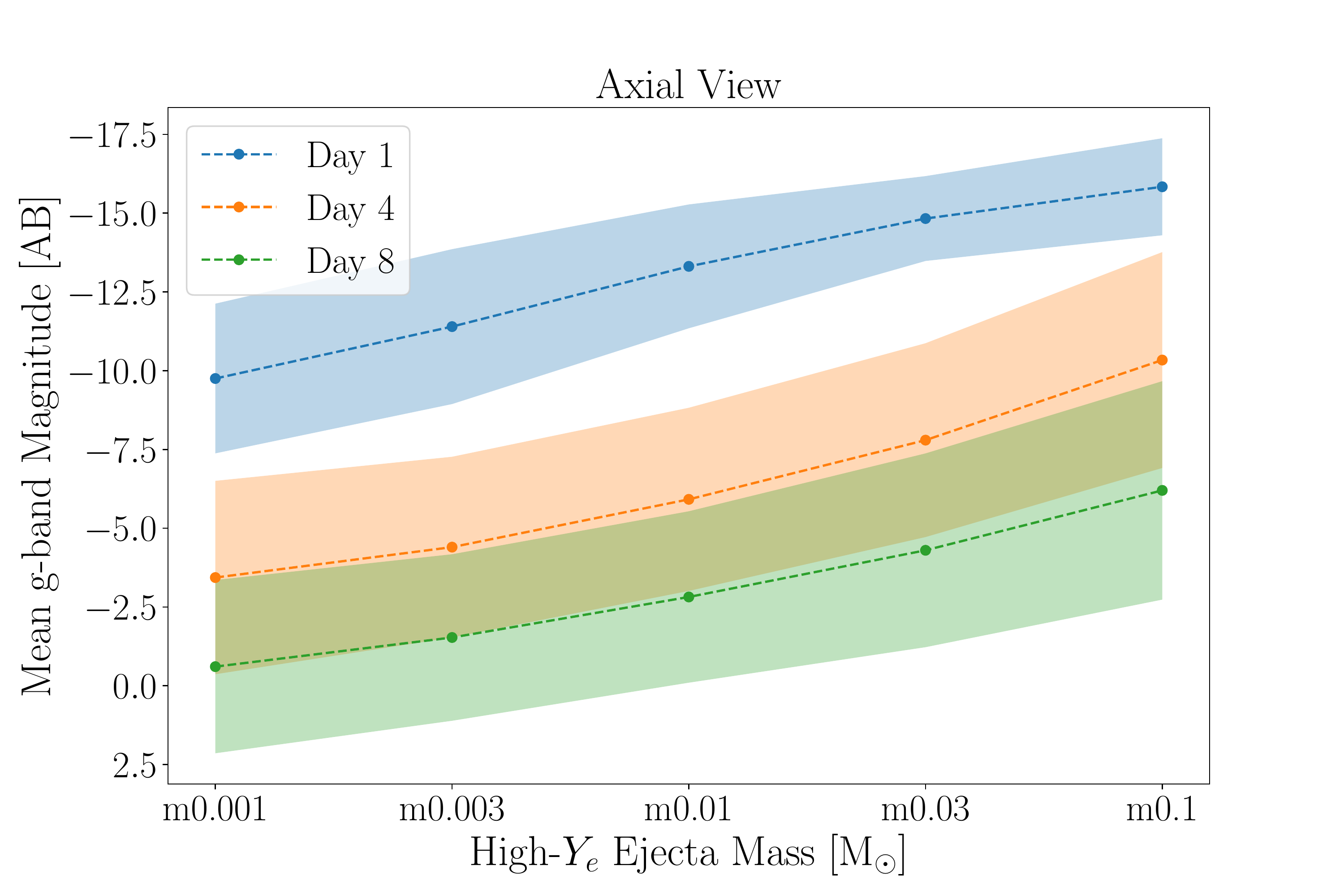}
  \includegraphics[width=0.9\columnwidth]{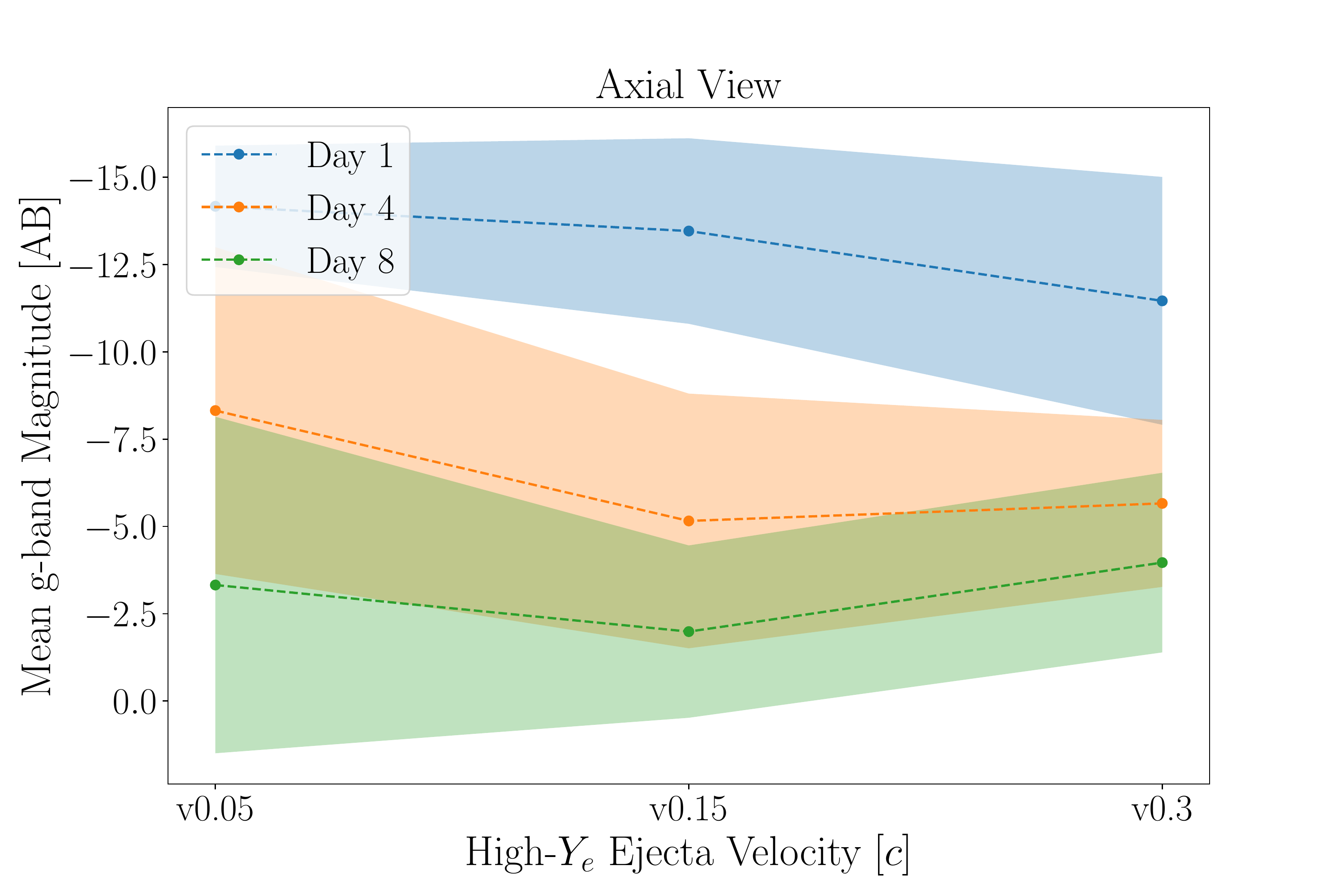}
  \caption{Mean $g$-band magnitude in axial view per low-$Y_e$ ejecta mass (top left)
    and velocity (top right), and high-$Y_e$ mass (bottom left) and velocity (bottom right),
    at days 1 (blue), 4 (orange), and 8 (green).
    The $g$-band magnitude trend is most apparent with increasing high-$Y_e$ mass,
    as expected.
    The mean $g$-band is not sensitive at day 1 to low-$Y_e$ ejecta mass, which is
    consistent with previous findings for early emission from 2-component models
    \citep{tanvir2017}.
  }
  \label{fg3:tren}
\end{figure*}

Figure~\ref{fg3:tren} has day 1, 4, and 8 axial $g$-band magnitudes versus low-$Y_e$
ejecta mass and velocity (top row), and high-$Y_e$ ejecta mass and velocity (bottom row).
In contrast to the bolometric luminosity, the standard deviation in the $g$-band
magnitude at day 1 is small compared to the mean magnitude.
However, this is partly a result of computing standard deviation in magnitudes.
As expected, we find that the variability of the $g$-band with respect
to the high-$Y_e$ mass is more pronounced than with respect to the low-$Y_e$ mass.
Similar to the bolometric luminosity, the mean $g$-band magnitude does not have
strong trends in either low- or high-$Y_e$ ejecta velocities.
The dependence of the $g$-band on the high-$Y_e$ velocity is more sensitive, but
apparently non-monotonic at days 4 and 8.
The drop in the mean from $0.05c$ to $0.15c$ is consistent with a reduced time
scale of the blue transient, and the increase in the mean from $0.15c$ to $0.3c$
is consistent with the high-$Y_e$ component reprocessing more emission from the
slower (or equal speed) low-$Y_e$ component.

\begin{figure*}
  \centering
  \includegraphics[width=0.9\columnwidth]{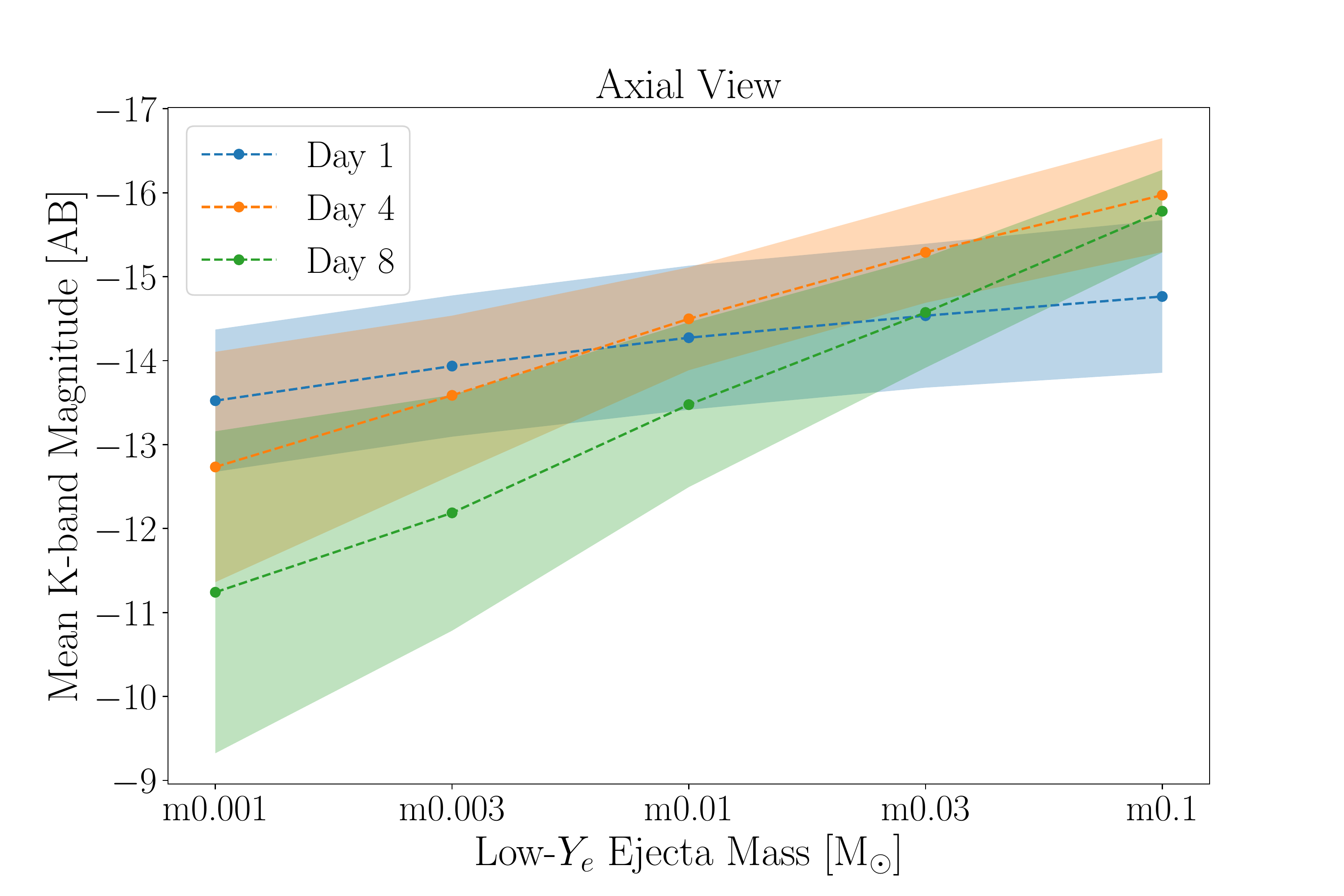}
  \includegraphics[width=0.9\columnwidth]{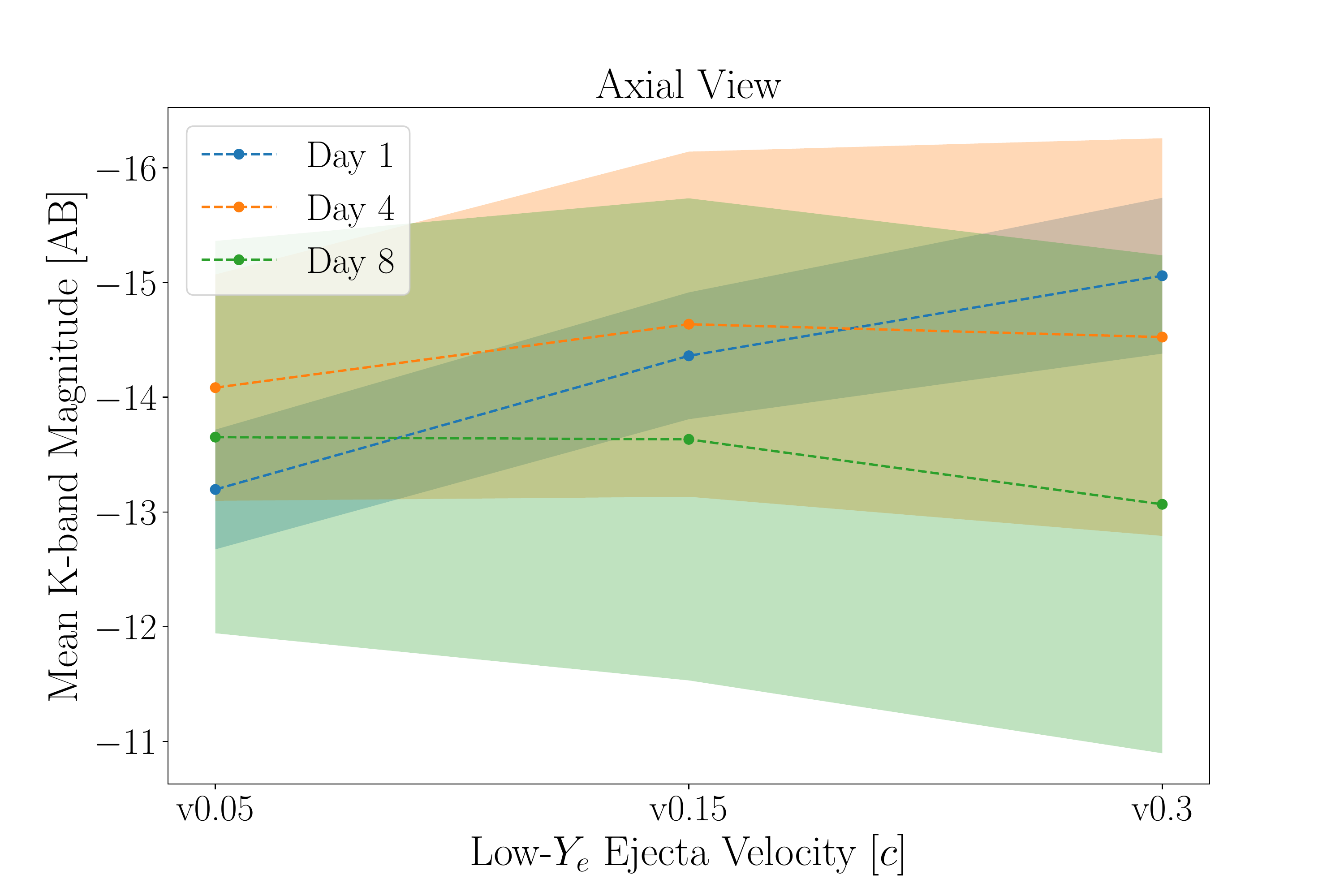} \\
  \includegraphics[width=0.9\columnwidth]{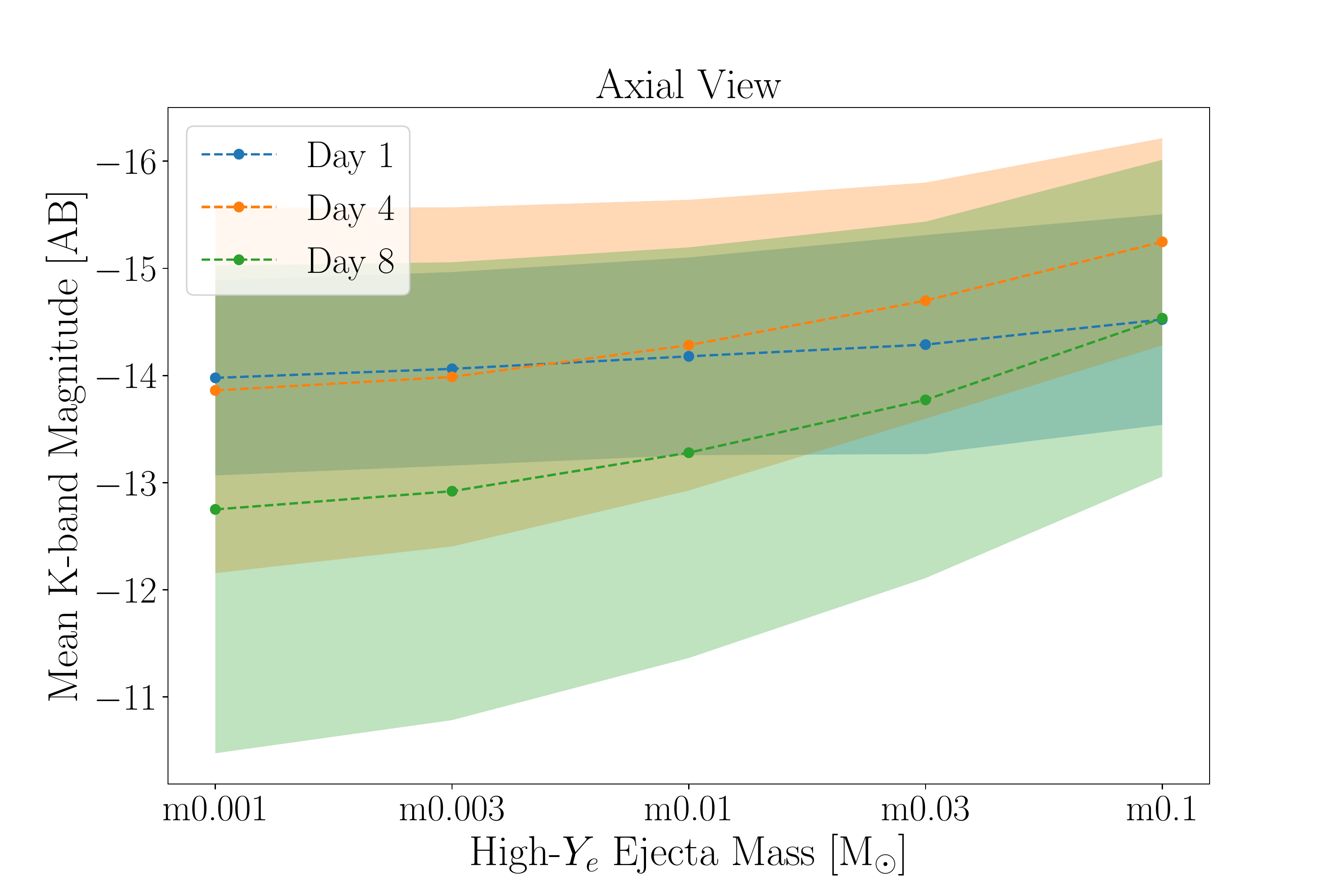}
  \includegraphics[width=0.9\columnwidth]{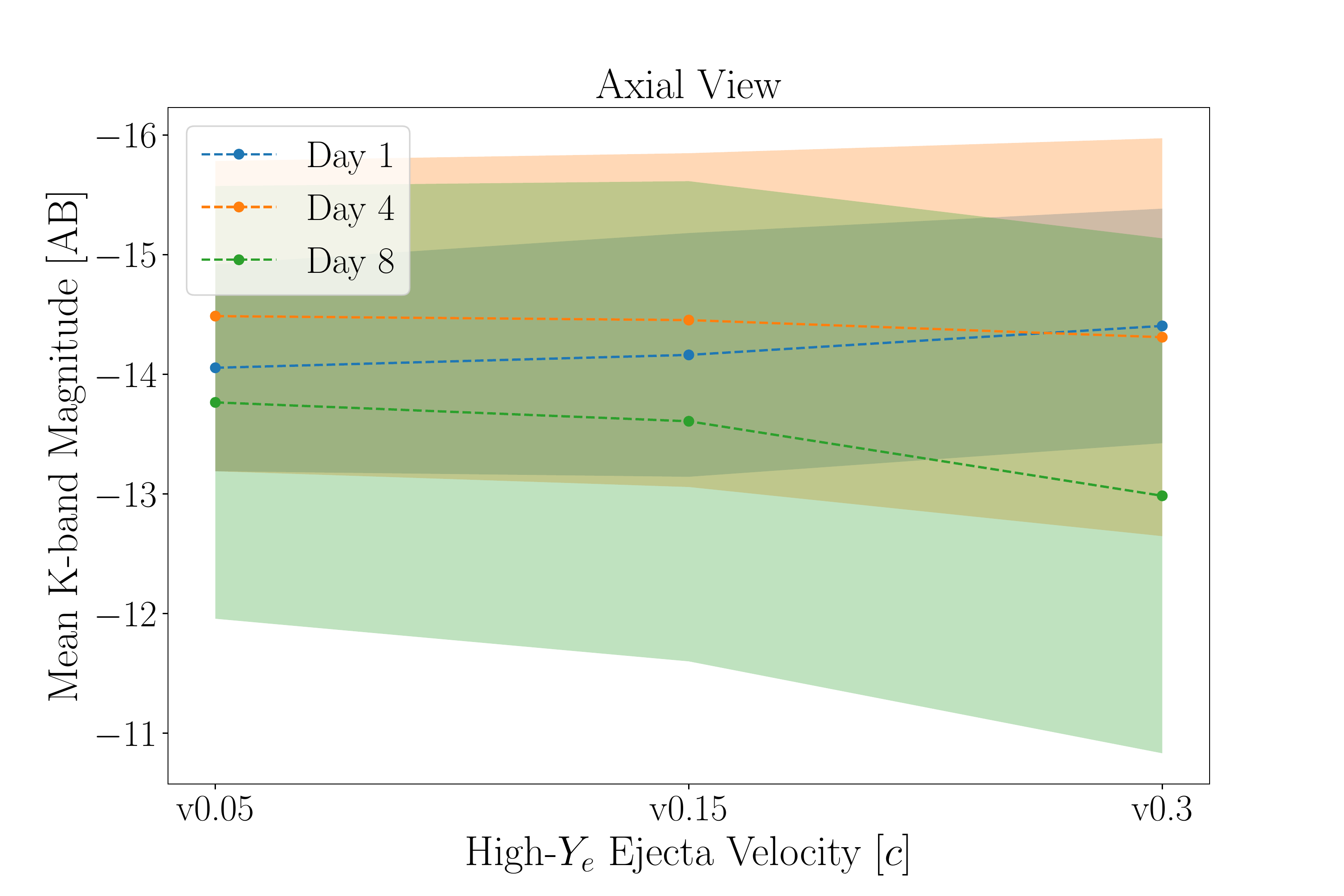}
  \caption{Mean $K$-band magnitude in axial view per low-$Y_e$ ejecta mass (top left)
    and velocity (top right), and high-$Y_e$ mass (bottom left) and velocity (bottom right),
    at days 1 (blue), 4 (orange), and 8 (green).
    The magnitude collectively trends upwards for both increasing low- and high-$Y_e$ ejecta.
    Moreover, the trend in low-$Y_e$ ejecta mass becomes significantly steeper at later time,
    indicating a growing impact of the low-$Y_e$ ejecta mass on $K$-band magnitude across
    models.
  }
  \label{fg4:tren}
\end{figure*}

Figure~\ref{fg4:tren} has the same data as Fig.~\ref{fg3:tren}, but for the much
redder $K$-band.
In contrast with the collective $g$-band results, the brightness in the $K$-band
is more sensitive to the low-$Y_e$ ejecta mass (top left).
Moreover, the change in mean magnitude between the grid mass extrema is larger for
later time (comparing the day 1, 4, and 8 trend curves).
This increase in sensitivity for later time is expected in the axial view, where
the high-$Y_e$ mass plays a more dominant role in setting the brightness across bands.
The magnitude of the $K$-band also trends upward for increasing high-$Y_e$ ejecta mass
(bottom left), though there is high variability across models for each mass value.

The collective trends of emission with respect to ejecta mass are consistent
with our expectations for the behavior of 2-component kilonova models.
It is more difficult to discern trends in velocity, but the weak trends that
do exist readily correspond to well-established explanation (for example, lanthanide
curtaining when the low-$Y_e$ component is faster than the high-$Y_e$ component).
However, similar to the diversity in angular variation between models, the large
standard deviations per fixed mass or velocity may obfuscate the identification
of other model properties.
In the following sections, we turn to the question of whether the two broadest
properties, the high-$Y_e$ morphology and composition, when partitioned into two
groups, can be statistically distinguished.

\subsection{High-$Y_e$ Morphology and Composition Populations}
\label{sec:wmcp}

\begin{figure}
  \centering
  \includegraphics[width=0.9\columnwidth]{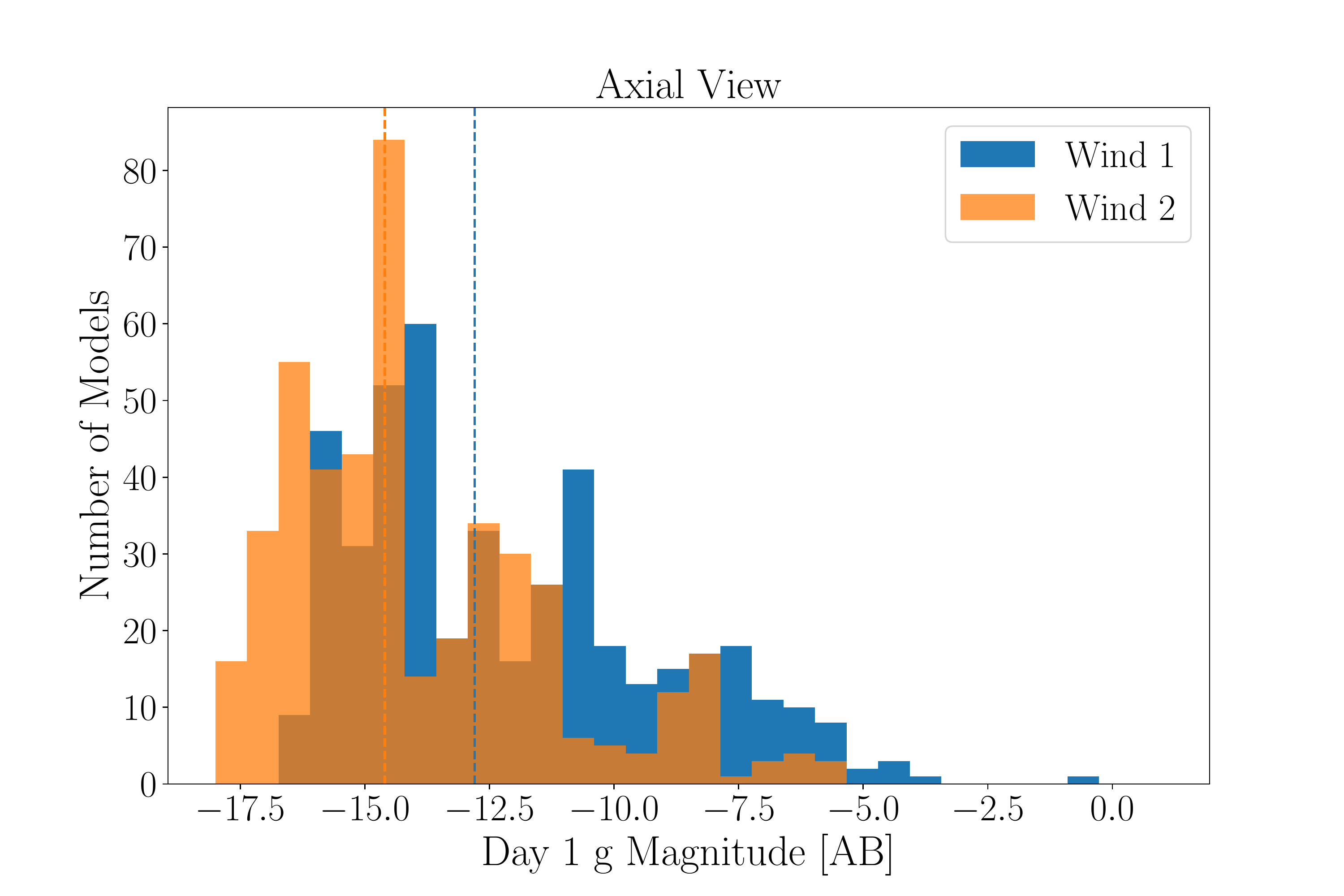} \\
  \includegraphics[width=0.9\columnwidth]{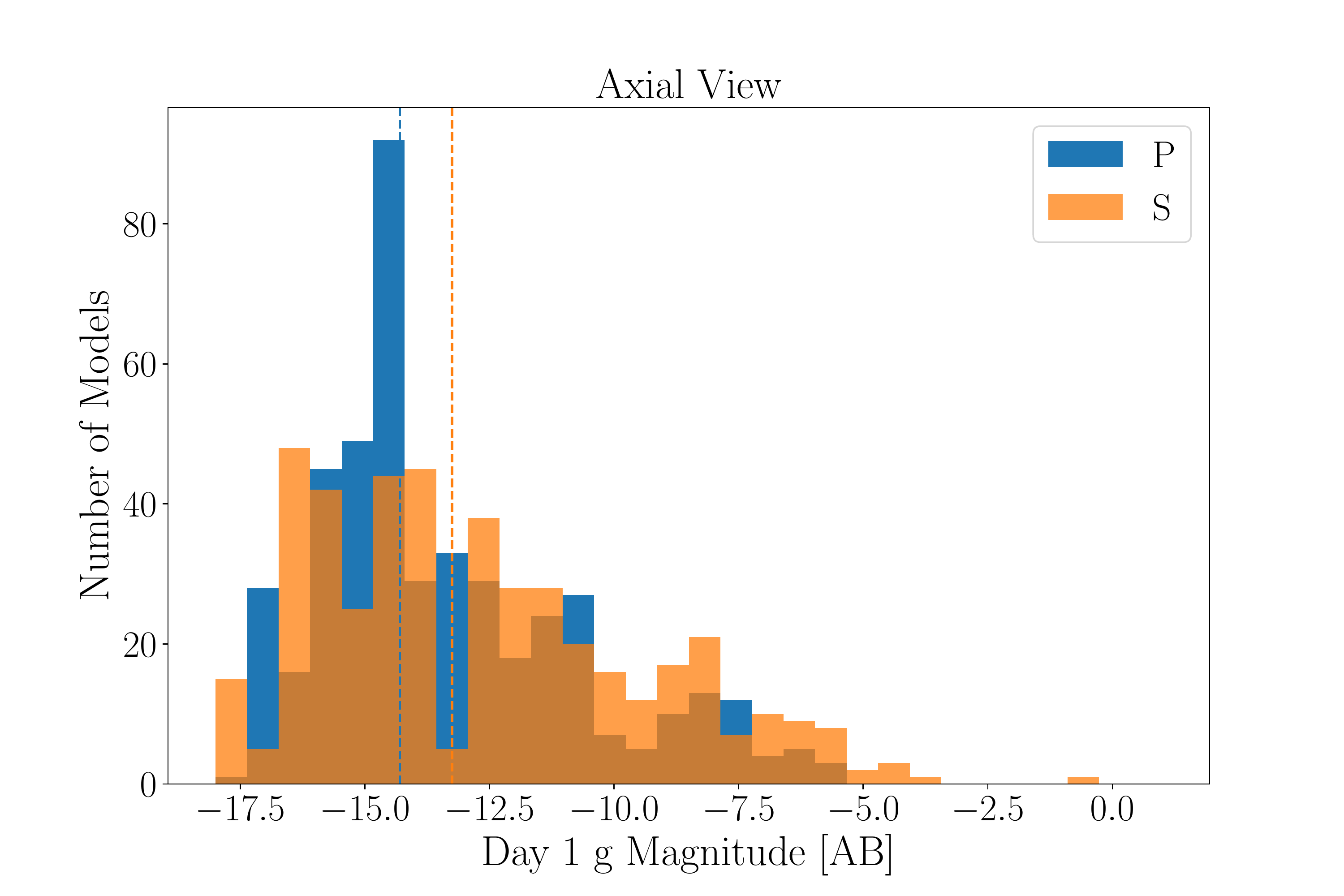}
  \caption{Histograms for fixed high-$Y_e$ composition (top) and morphology (bottom)
    of axial $g$-band absolute magnitudes over 30 uniformly spaced bins
    between magnitude values of -18 and 1.
    The histograms indicate a systematic shift between composition distributions.
    Medians are shown as dashed vertical lines colored by distribution.}
  \label{fg1:wmcp}
\end{figure}

We may test a null hypothesis that the data belong to the same distribution
(i.e. same mean and standard deviation) when partitioned into groupings by model
properties.
In particular, given the differing degrees of lanthanide curtaining shown
Section~\ref{sec:vang} and the amount of variance in trends with mass and velocity
shown in Section~\ref{sec:tren}, one may ask if broader model properties,
such as high-$Y_e$ ejecta composition or morphology, can be statistically distinguished.
Comparing these broad groupings could be performed with an analysis of variance,
or similar statistical technique, to determine if the groups are distinguishable
by their observable photometric properties (luminosity, magnitudes), assuming a fixed viewing angle
for all models.
However, with the model set, we may not be able to assume the observables are
normally distributed; e.g. we may not be able to assume that the frequency of
luminosity in a particular range follows a normal distribution over luminosity bins.
Considering Fig.~\ref{fg1:wmcp}, we see that the distribution of axial
absolute g-band magnitudes do not apparently follow normal distributions.
While it may be possible to transform the observational data points to get more
normal distributions, we may instead use some simple non-parametric statistical
analysis to distinguish data groups.

In the following sub-sections, we explore the application of the non-parametric
Mann-Whitney U test to find a subset (or ``sub-vector'') of data from each model
that can distinguish morphology or composition.
With the sub-vector of data, we then perform a logistic regression over the
models, and apply the resulting fitting parameters to fit the model data itself.
With this series of calculations, we intend to demonstrate that the high-$Y_e$
composition of these models is easier to categorize into distinct groups than
the high-$Y_e$ ejecta morphology.

\subsubsection{Mann-Whitney U Test of Magnitudes and Luminosity}
\label{sec:mwut}

If the distributions within the partitioned groups are comparable in form,
we may apply the non-parametric Mann-Whitney U test
which, for two groups, tallies the number of times ($U$) that the samples in one
group precede (in some ordering) the samples in another~\citep{mann1947},
\begin{equation}
  U = mn + \frac{m(m+1)}{2} - T \;\;,
  \label{eq1:mwut}
\end{equation}
where $m$ and $n$ are the number of data points in each group, and $T$ is the
sum of the ranks of the $m$ elements from one group in the $n+m$ ordered ranking
of the total data set (the Wilcoxon statistic).
The ``ranks'' are simply positions from sorting in the $n+m$ size list of all
the data, which implies
\begin{multline}
  \sum_i^nr_i + \sum_j^ms_j = \sum_i^nr_i + T \\
  = \sum_k^{n+m}k = nm + \frac{m(m+1)}{2} + \frac{n(n+1)}{2} \;\;,
  \label{eq2:mwut}
\end{multline}
where $r_i$ and $s_j$ are rank subsets of $\{1,\ldots, m+n\}$ of the size $n$
and $m$ groups, respectively.
The minimal rank-sum of the $r_i$ values is $n(n+1)/2$ (just the triangular
number), which corresponds to a $U$ value of 0.
An arbitrary ranking (with some $r_i > s_j$) can be formed from the $U=0$
ranking by first permuting $r_n=n$ with $s_j=n+j$, then $r_{n-1}=n-1$ with
$s_{j'}=n+j'$ where $j'<j$, and so forth.
From this procedure, each permutation must increase the $U$ statistic by the
difference of the old and new value for each $r_i$, since only values of $s_j$
will be passed over when $r_i$ is permuted.
Consequently,
\begin{equation}
  U = \sum_i^n(r_i-r_i') = \sum_i^nr_i - \sum_i^nr_i'
  = \sum_i^nr_i - \frac{n(n+1)}{2} \;\;,
  \label{eq3:mwut}
\end{equation}
where $r_i'$ are the old ranks for $U=0$.
Substituting Eq.~\eqref{eq3:mwut} into Eq.~\eqref{eq2:mwut} gives
Eq.~\eqref{eq1:mwut}.
The probability of a particular $U$ value can be found by counting the
number of $n+m$ rank sequences that give $U$ from Eq.~\eqref{eq1:mwut},
and can be computed with a recurrence relation (see~\citet{mann1947}).
We use the SciPy stats package~\citep{virtanen2020}, which has the
Mann-Whitney U test as an intrinsic function.

The $U$ value can be used to test the null hypothesis that the cumulative
distribution functions describing the two data sets are the same.
In Fig.~\ref{fg1:wmcp}, the histograms are grouped by high-$Y_e$ ejecta composition
(top) and high-$Y_e$ ejecta morphology. In these figures, it can be seen that the
probability distributions are similarly
shaped but may be significantly shifted in g-band brightness (especially for the
grouping by composition).
The distributions for the $K$ band and bolometric luminosities are also similar
for these partitions of the data.
Consequently, the conditions evidently suffice to use the Mann-Whitney U test
over the data in the expanded form of Table~\ref{tb1:app}.

\begin{figure}
  \centering
  \includegraphics[width=0.9\columnwidth]{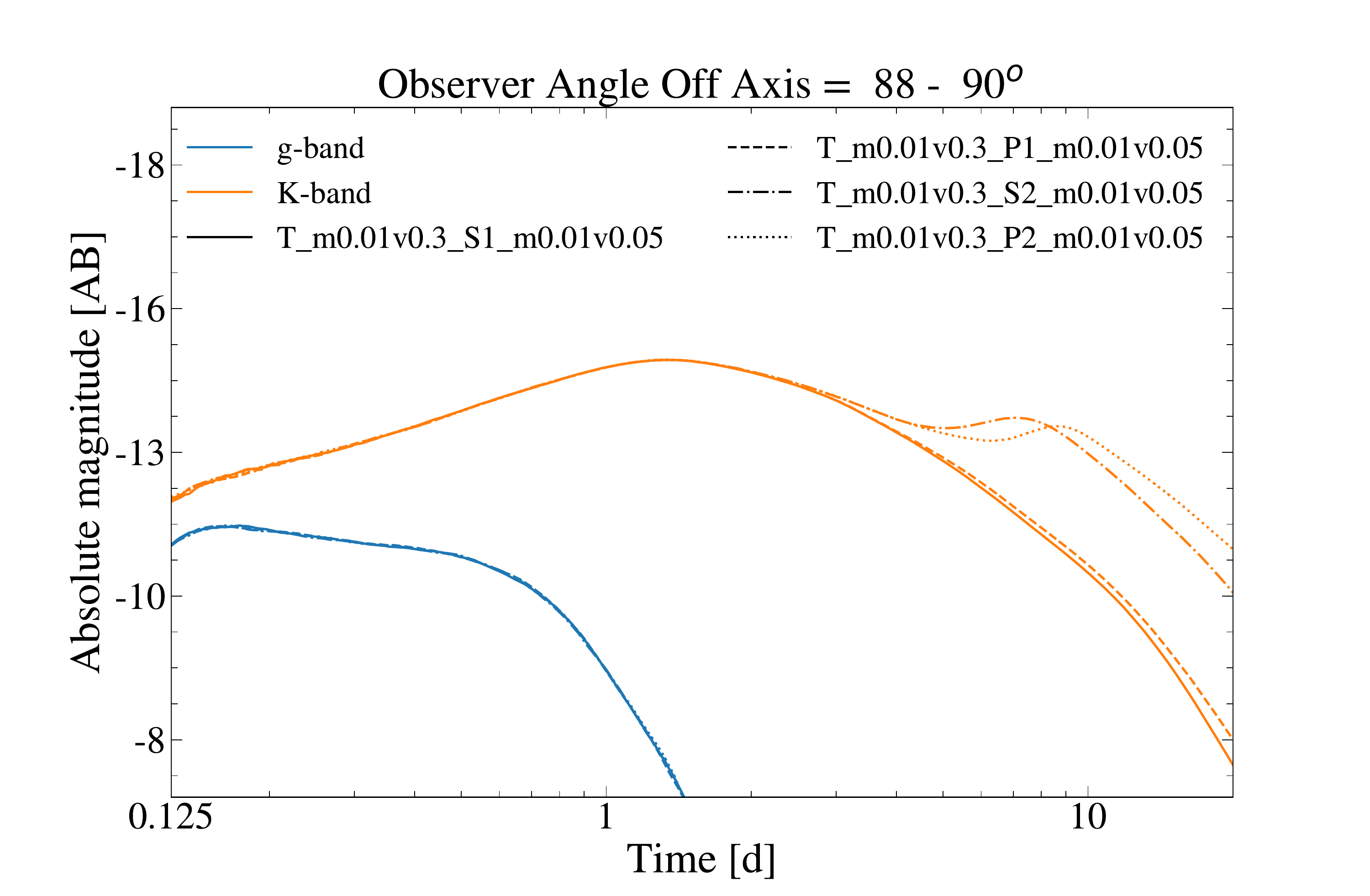}\\
  \includegraphics[width=0.9\columnwidth]{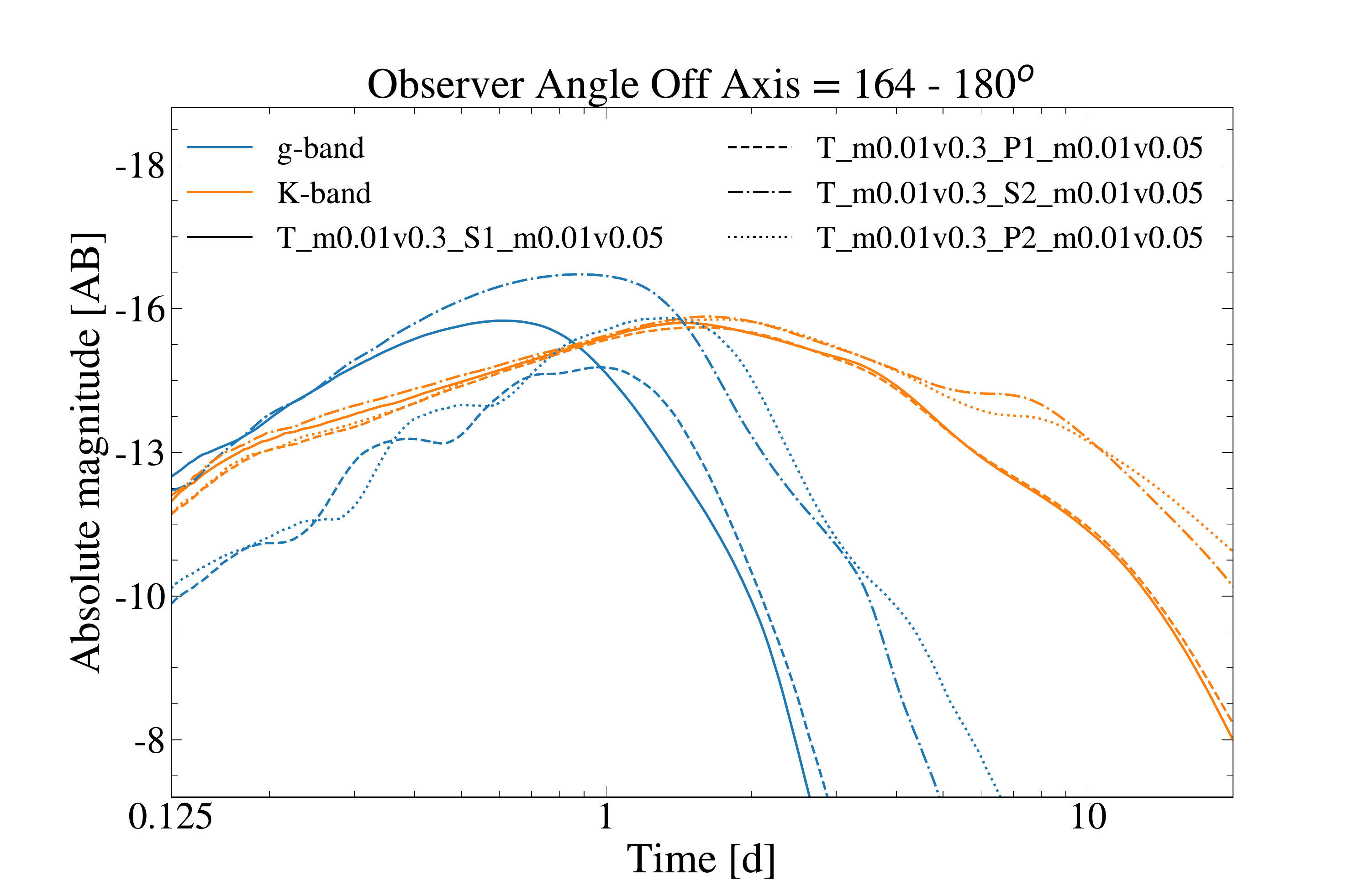}
  \caption{g (blue) and K (orange) band absolute AB magnitudes for models
    with $0.01$ M$_{\odot}$ mass for both high- and low-$Y_e$ ejecta, $0.3c$
    low-$Y_e$ ejecta speed, and $0.05c$ high-$Y_e$ ejecta speed.
    The high-$Y_e$ ejecta are spherical ("S", wind 1 = solid, wind 2 =
    dot-dashed) and lobed ("P", wind 1 = dashed, wind 2 = dotted).
    The top panel is for the edge-on view and the bottom panel is for an
    axial view; as expected, the $g$-band is several magnitudes brighter
    in the axial view.
    }
  \label{fg1:rsts}
\end{figure}

Tables~\ref{tb1:wmcp} and~\ref{tb2:wmcp} show results of the Mann-Whitney U test
for the high-$Y_e$ morphology and composition groups, respectively.
These results include the $U$ statistic, $p$-value, and common language effect
size $f$,
which is the fraction of all possible pairs of data from each group that support
an alternative to the null hypothesis (the null hypothesis is supported by
values of $f$ close to 0.5, see~\citet{mcgraw1992}).
Shaded table cells indicate a $p$-value less than 0.05 (or 5\%), a standard
significance level for rejecting the null hypothesis.
For the morphology test, the $p$-values indicate that the most significantly
distinguishable cumulative distribution functions are over the $g$-band magnitudes,
with axial views providing higher significance at the later days.
This is consistent with the $g$-band magnitude trends shown in Fig.~\ref{fg1:rsts}
for the on-axis view (lower panel): by day 4 there is a notable difference in
the P and S morphologies for the wind 2 composition.
Thus the Mann-Whitney U test verifies the apparent systematic shift between the
distributions for each composition shown in Fig.~\ref{fg1:wmcp}.

In contrast, the $p$-values for the $K$-band magnitude increase with time in both
axial and edge views of the ejecta, indicating that the morphology distributions
over the $K$ band become less statistically distinguishable as time progresses.
The sample $K$-band magnitude curves in Fig.~\ref{fg1:rsts} provide evidence for
this result as well, where the late time curves group by morphology and separate
by composition.
Moreover, only one of the six tabulated values is significant at the 5\% level,
while the other five do not reject the null hypothesis at this level.
The bolometric luminosity likewise only has one $p$-value less than 0.05, for the
day 1 edge view of the ejecta.

\begin{table*}
  \centering
  \caption{Results of Mann-Whitney U test of high-$Y_e$ ejecta morphology (S group
    versus P group) per axial and
    edge views of $g$ band, $K$ band and bolometric luminosity $L_{\rm bol}$: $U$ statistic, $p$ value,
    and common language effect size $f$ at days 1, 4, and 8.
    $p$-values less than 5\% are highlighted.}
  \begin{tabular}{lc|ccc|ccc|ccc}
    \hline\hline
    & & \multicolumn{3}{|c}{g} & \multicolumn{3}{|c}{K} & \multicolumn{3}{|c}{$L_{\rm bol}$} \\
    & Day & $U$ & $p$ & $f$ & $U$ & $p$ & $f$ & $U$ & $p$ & $f$ \\
    \hline
    & 1 & 90486 & \cellcolor{blue!15} $5.77\times10^{-3}$ & 0.447 & 116240 & \cellcolor{blue!15} $1.20\times10^{-4}$ & 0.574 & 96588 & 0.232 & 0.477 \\
    Axial & 4 & 78367 & \cellcolor{blue!15} $4.39\times10^{-9}$ & 0.387 & 107518 & 0.108 & 0.531 & 99476 & 0.649 & 0.491 \\
    & 8 & 76170.5 & \cellcolor{blue!15} $1.26\times10^{-10}$ & 0.376 & 100989.0 & 0.947 & 0.499 & 103208 & 0.615 & 0.510 \\
    \hline
    & 1 & 83241.5 & \cellcolor{blue!15} $3.86\times10^{-6}$ & 0.411 & 107800.5 & 0.0928 & 0.532 & 109573 & \cellcolor{blue!15} 0.0328 & 0.541 \\
    Edge & 4 & 94800.5 & 0.098 & 0.468 & 104131.5 & 0.460 & 0.514 & 103292 & 0.601 & 0.510 \\
    & 8 & 109313 & \cellcolor{blue!15} 0.0386 & 0.540 & 97343.5 & 0.316 & 0.481 & 106616 & 0.168 & 0.526 \\
    \hline
  \end{tabular}
  \label{tb1:wmcp}
\end{table*}

The Mann-Whitney U test with partitioning in groups by composition produces
more significant differences in cumulative distributions than morphology: 16
significant distribution comparisons compared to 7 for morphology at the 5\% level.
The $p$-value for the $g$ band is effectively zero for both axial and edge views
of the ejecta (similar to the morphology test along the axial view, the $p$-value
decreases towards later time).
In contrast to the morphology grouping, the $K$ band at late time is
a strong indicator of composition in the high-$Y_e$ component.
The $K$-band magnitudes are close at day 1, but these magnitudes become
systematically lower for the wind 1 composition; this trend is reflected
in the lower $p$-values at later time.
The final noteworthy difference with the morphology test is in the bolometric
luminosity: for the composition test it is a significant indicator in the
difference between the two populations.
The difficulty of the Mann-Whitney U test in distinguishing morphology, relative
to composition, is consistent with the morphology distributions being in closer
alignment in Fig.~\ref{fg1:wmcp}.
This pattern is noteworthy given the impact that morphology can have in brightness
\citep{korobkin2020}.

However, the Mann-Whitney U test (as presented here) does not test for differences
in correlations among observables.
For instance, the covariance of luminosity and $K$-band magnitude might be stronger
for one morphology (composition) than the other, which would be a distinguishing factor
between the morphology (composition) groups.
We can calculate the Pearson correlation coefficient~\citep{benesty2009} amongst the 9
data values per model per viewing angle ($g$ band, $K$ band, bolometric luminosity at
day 1, 4, 8 for 36 comparisons per model, comparing axial-to-axial, axial-to-edge,
or edge-to-edge views),
\begin{equation}
  \varrho_{i,j}^{(P,v,w)} =
  \frac{\sum_{k=1}^{450}(o_{i,k}^{(P,v)}-\bar{o}_i^{(P,v)})(o_{j,k}^{(P,w)}-\bar{o}_j^{(P,w)})}
       {\sqrt{\sum_{k=1}^{450}(o_{i,k}^{(P,v)}-\bar{o}_i^{(P,v)})^2}
         \sqrt{\sum_{k=1}^{450}(o_{j,k}^{(P,w)}-\bar{o}_j^{(P,w)})^2}} \;\;,
       \label{eq4:mwut}
\end{equation}
where $i$ and $j$ are indices going from 1 to 9 labeling each observable, $o_{i,k}^{(P,v)}$ is the
value of the observable $i$ for model with morphology P indexed at $k$, in viewing angle indexed $v$;
so $\varrho_{i,j}^{(P,v,w)}\in[-1, 1]$ is the Pearson correlation coefficient between $i$ and $j$ between
viewing angles $v,w\in\{\text{axial},\text{edge}\}$.
The maximum relative difference of correlation coefficients between morphologies is calculated as
\begin{equation}
  \eta^{(P,S;v,w)} = \max_{i,j}\left\{\frac{|\varrho_{i,j}^{(P,v,w)} - \varrho_{i,j}^{(S,v,w)}|}
      {\min\{|\varrho_{i,j}^{(P,v,w)}|, |\varrho_{i,j}^{(S,v,w)}|\}}\right\} \;\;.
      \label{eq5:mwut}
\end{equation}
Equation~\eqref{eq5:mwut} applies to composition if P is replaced by ``wind 1'' and S is replaced by
``wind 2''.
Using Eq.~\eqref{eq5:mwut} for either morphology groups or composition groups, we find that the
overall maximum relative difference in Pearson correlation coefficients occurs for morphology in the
correlation between day 1 bolometric luminosity in the edge view and day 8 bolometric luminosity in
the axial view: the correlations have the same sign ($\varrho_{i,j}^{(P,\text{axial},\text{edge})}=0.71$
and $\varrho_{i,j}^{(S,\text{axial},\text{edge})}=0.86$), but is stronger for S morphology by
$\eta^{(P,S;\text{axial},\text{edge})}\sim$21\%.
For comparison the largest relative difference in correlation between composition groups is
$\eta^{(1,2;\text{edge},\text{edge})}\sim$15\%,
for the correlation between day 1 $K$-band magnitude in the edge view and day 8 $g$-band magnitude also
in the edge view.
The Pearson correlation coefficients suggest that the impact of morphology on correlations between
observables in different viewing angles per model can further distinguish the morphology groups.
Given an observation is viewed at only one angle from Earth, the utility of the difference in the
multi-angle correlations between the morphology groups in comparing an observation to the model grid
is not readily apparent.
Consequently, we focus on using the results of the Mann-Whitney U-test in forming a grid-observation
comparison in the subsequent sections.

\begin{table*}
  \centering
  \caption{Results of Mann-Whitney U test of high-$Y_e$ ejecta composition (wind 1
    group versus wind 2 group)
    per axial and edge views of $g$ band, $K$ band and bolometric luminosity $L_{\rm bol}$:
    $U$ statistic, $p$ value, and common language effect size $f$ at days 1, 4, and 8.
    $p$-values less than 5\% are highlighted.}
  \begin{tabular}{lc|ccc|ccc|ccc}
    \hline\hline
    & & \multicolumn{3}{|c}{g} & \multicolumn{3}{|c}{K} & \multicolumn{3}{|c}{$L_{\rm bol}$} \\
    & Day & $U$ & $p$ & $f$ & $U$ & $p$ & $f$ & $U$ & $p$ & $f$ \\
    \hline
    & 1 & 138996.5 & \cellcolor{blue!15} $3.60\times10^{-22}$ & 0.686 & 105100 & 0.323 & 0.519 & 70296.5 &  \cellcolor{blue!15} $2.05\times10^{-15}$ & 0.347 \\
    Axial & 4 & 145124.5 & \cellcolor{blue!15} $2.247\times10^{-29}$ & 0.717 & 115642 &  \cellcolor{blue!15} $2.23\times10^{-4}$ & 0.571 & 73292.5 &  \cellcolor{blue!15} $7.50\times10^{-13}$ & 0.362 \\
    & 8 & 158541.5 & \cellcolor{blue!15} $7.04\times10^{-49}$ & 0.783 & 123396.5 &  \cellcolor{blue!15} $1.34\times10^{-8}$ & 0.609 & 72340.5 &  \cellcolor{blue!15} $1.19\times10^{-13}$ & 0.357 \\
    \hline
    & 1 & 125232.5 & \cellcolor{blue!15} $7.69\times10^{-10}$ & 0.618 & 101830.5 & 0.882 & 0.503 & 82928.5 & \cellcolor{blue!15} $2.62\times10^{-6}$ & 0.409 \\
    Edge & 4 & 130887 & \cellcolor{blue!15} $2.94\times10^{-14}$ & 0.646 & 114047.5 & \cellcolor{blue!15} $1.03\times10^{-3}$ & 0.563 & 77066.5 & \cellcolor{blue!15} $5.55\times10^{-10}$ & 0.381 \\
    & 8 & 141209.5 & \cellcolor{blue!15} $1.17\times10^{-24}$ & 0.697 & 126966.5 & \cellcolor{blue!15} $4.21\times10^{-11}$ & 0.627 & 70454 & \cellcolor{blue!15} $2.69\times10^{-15}$ & 0.348 \\
    \hline
  \end{tabular}
  \label{tb2:wmcp}
\end{table*}

Finally, we note that we have verified the significant variables in Tables~\ref{tb1:wmcp} and~\ref{tb2:wmcp}
with the Kolmogorov-Smirnov test, which also tests if two populations belong to the same distribution,
but uses a distance metric between the partitioned distributions~\citep{smirnov1948}.
In Section~\ref{sec:irls}, we attempt to use the Mann-Whitney U test results to
inform a logistic regression.
The intent is to determine the effectiveness of this logistic regression in
observable-based categorization of models into the morphology or composition groups.

\subsubsection{Iteratively Reweighted Least Squares Categorization}
\label{sec:irls}

The Mann-Whitney U test is a method for assessing whether two groups of data belong
to one distribution over a particular parameter (the null hypothesis), but does not
readily provide a way to categorize new data into a group.
Consequently, for an observation or model that is not in the grid of 900, we need
a different technique to determine the probability that it belongs to a particular
category (for instance, of composition or morphology).
One simple approach is to examine the data with the most significant $p$-values
in Tables~\ref{tb1:wmcp} or~\ref{tb2:wmcp}, and perform a multivariate logistic
regression on those values.
Specifically, for each model we use the data corresponding to the shaded $p$-values
from those tables to perform the regression, which depends on the model property:
\begin{enumerate}
\item the $g$ band and day 1 $K$ band for morphology,
\item and all data except day 1 $K$ band for composition.
\end{enumerate}
The fitted logistic can then be used to calculate a probability of a new observation's
inclusion in one of the groups.
We attempt to fit the logistic regression parameters using an Iteratively
Reweighted Least Squares (IRLS) procedure~\citep{holland1977}, which is
straightforward to implement using matrix manipulations available in NumPy's linear
algebra package~\citep{harris2020}.
We concatenate size-$m$ sub-vectors $\vec{y}_U$ for each model into a
900-by-$m$ matrix $\mathbf{Y}_U$, where $m$ is the number of variables over which the
Mann-Whitney U test distributions are significantly different.
A logistic curve evaluated at each of the 900 data is
\begin{equation}
  \label{eq1:irls}
  \vec{\mu} = \frac{1}{1 + e^{-w_0-\mathbf{Y}_U\cdot\vec{w}}} \;\;,
\end{equation}
where $w_0$ and $\vec{w}$ are a scalar and size-$m$ vector of weights.
The IRLS procedure numerically solves for these weights with the following steps:
\begin{enumerate}
\item Calculate
  \begin{equation}
    \label{eq2:irls}
    \mathbf{M}_0^{(k)} = \bms{\mu}^{(k)}\cdot(\mathbf{I}-\bms{\mu}^{(k)}) \;\;,
  \end{equation}
  where superscript $(k)$ indicates evaluation at the $k^{\mathrm{th}}$ iteration,
  $\mathbf{M}_0^{(k)}$ is a 900x900 matrix, $\bms{\mu}^{(k)}$ is a 900x900
  diagonal matrix formed from the vector in Eq.~\eqref{eq1:irls}, and
  $\mathbf{I}$ is a 900x900 identity matrix.
\item Calculate
  \begin{equation}
    \label{eq3:irls}
    \mathbf{M}_1^{(k)}
    = (\mathbf{Y}_U^{(k)})^T\cdot\mathbf{M}_0^{(k)}\cdot\mathbf{Y}_U^{(k)} \;\;,
  \end{equation}
  where superscript $(k)$ indicates evaluation at the $k$th iteration,
  $\mathbf{M}_1^{(k)}$ is an $m+1\times m+1$ matrix, and superscript $T$ indicates
  a matrix transpose.
\item Calculate
  \begin{equation}
    \label{eq4:irls}
    \vec{M}_2^{(k)} = (\mathbf{Y}_U^{(k)})^T\cdot\left(
    \mathbf{M}_0^{(k)}\cdot\mathbf{Y}_U^{(k)}\cdot(w_0^{(k)}, \vec{w}^{(k)})^T
    + \vec{y} - \vec{\mu}^{(k)}\right) \;\;,
  \end{equation}
  where $\vec{M}_2^{(k)}$ is a vector of size $m+1$ and $\vec{y}$ is a vector
  of 450 one's and 450 zero's, where a one indicates an S morphology or wind 2
  composition.
\item Evaluate the weights for the next iteration,
  \begin{equation}
    \label{eq5:irls}
    (w_0^{(k+1)}, \vec{w}^{(k+1)})
    = (\mathbf{M}_1^{(k)})^{-1}\cdot\vec{M}_2^{(k)} \;\;,
  \end{equation}
  where superscript -1 indicates a matrix inverse.
\end{enumerate}
The above steps are similar to a multivariate Newton iteration and maximize
the log-likelihood of the logistic in Eq.~\eqref{eq1:irls} relative to the
vector of true categorization, $\vec{y}$.
The computational expense for the matrix inversion is mitigated by restricting
$m$ to be the number of significant data points from the Mann-Whitney U test.
In our calculation, the weights tend to be converged after five iterations.
We constrain the number of parameters to four for the morphology test (three $g$ band
and one $K$ band point) and eight for the composition test (three $g$ band, two $K$ band,
and three bolometric luminosity points).
This choice implies five and nine entries for $(w_0,\vec{w})$ for the morphology and
composition tests, respectively.

\begin{figure}
  \centering
  \includegraphics[width=0.9\columnwidth]{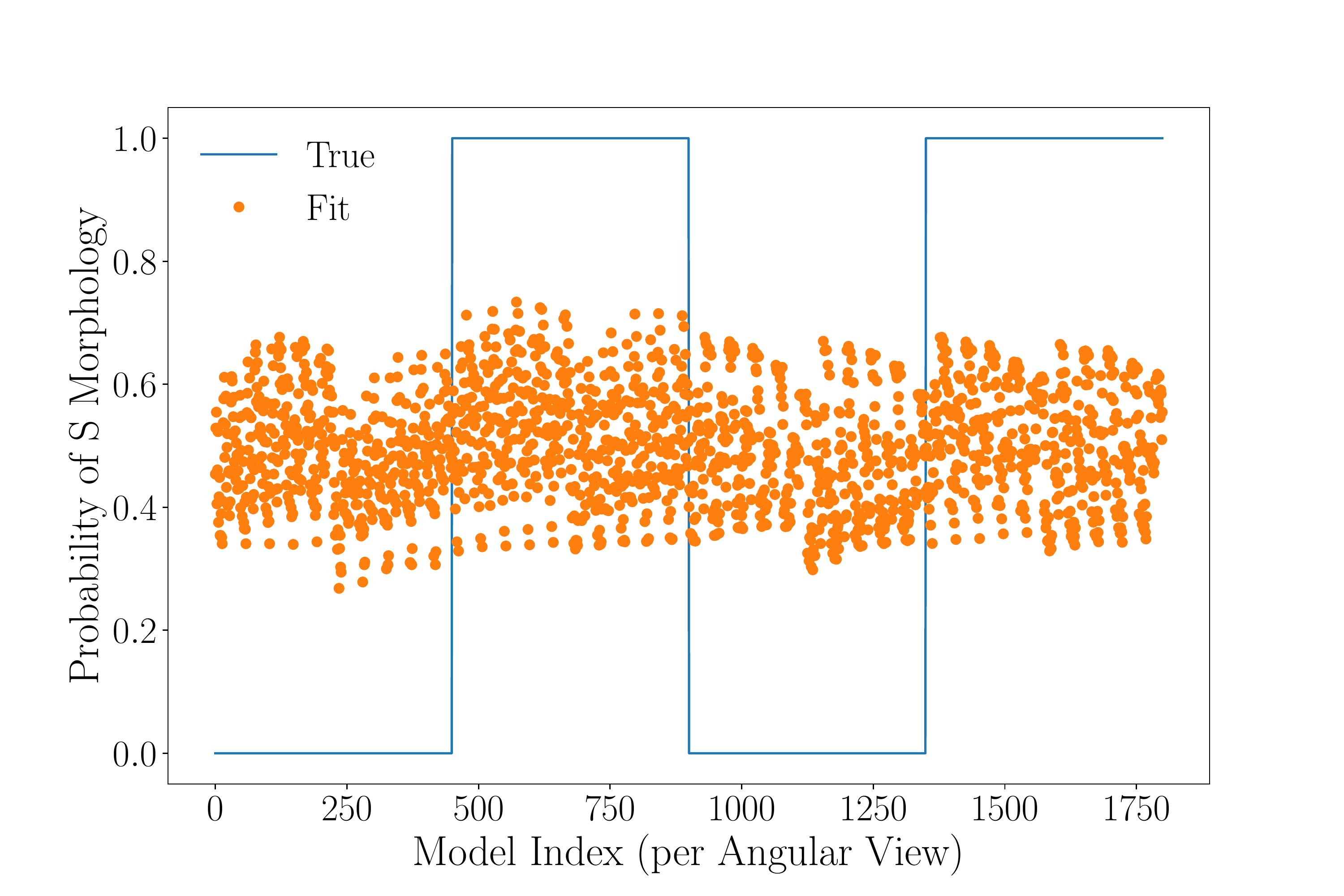} \\
  \includegraphics[width=0.9\columnwidth]{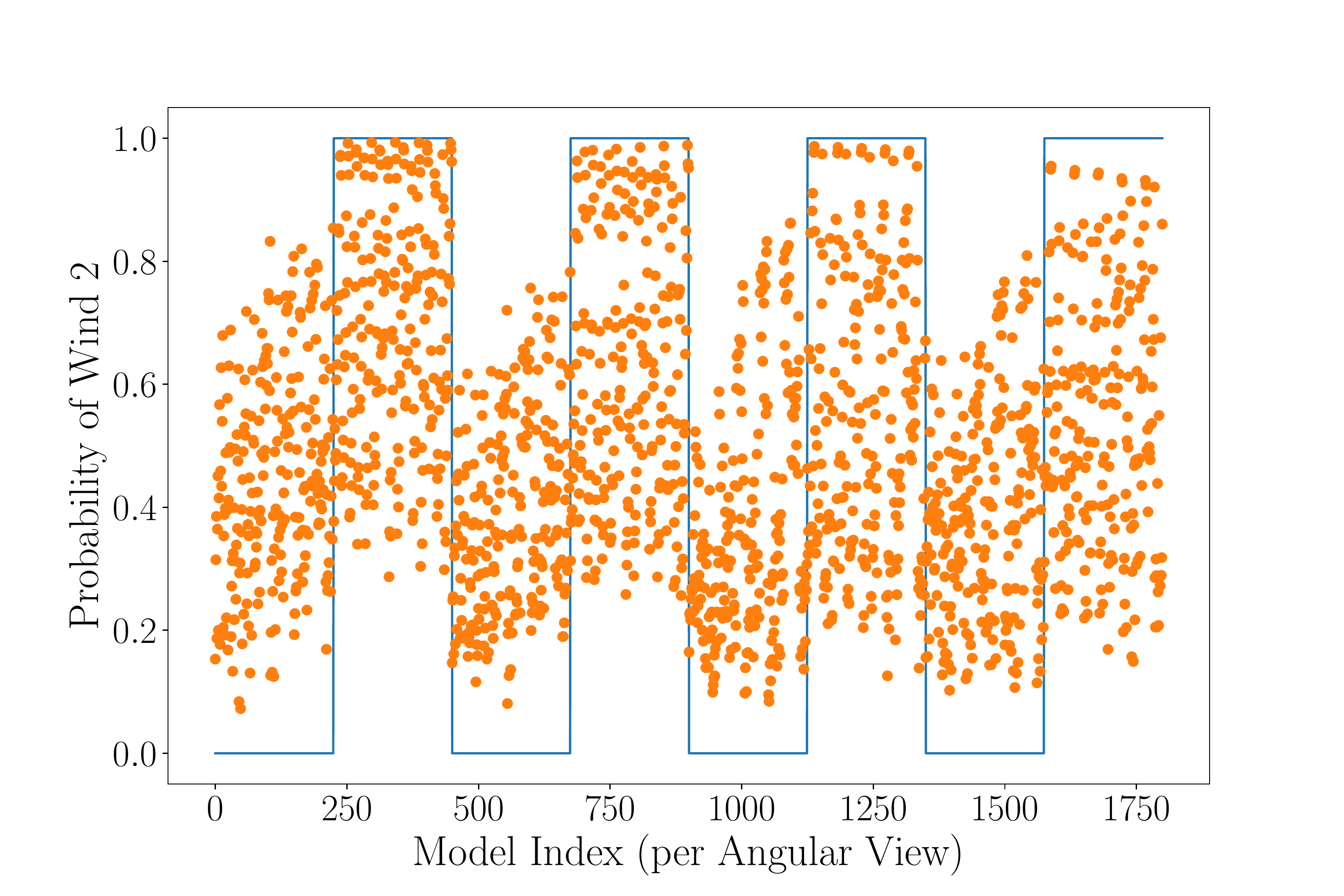}
  \caption{
    Fitted probabilities (orange, dotted) of the high-$Y_e$ component of each model
    having S morphology (top) or wind 2 composition (bottom) versus a model enumeration.
    Also shown are the true probabilities (0\% or 100\%) of inclusion in a group
    (blue, solid).
  }
  \label{fg2:wmcp}
\end{figure}

Figure~\ref{fg2:wmcp} displays the probability of group inclusion versus model index (doubled over
axial and edge views, with edge views indexed 900-1799) for the morphology test (top) and
for the composition test (bottom).
We have used the data set used in the regression to obtain logistic parameters, and have
used this logistic fit to estimate the probability that each model is included in its
own group.
Despite using the parameters producing the lowest $p$-values in the Mann-Whitney U test
for morphology, the logistic regression of the morphology does not fit the true
categorization, and hence does not provide predictive capability for new models (within
the scope of the limited data per model being used).
The logistic regression for the composition groups fares somewhat better, as can be seen
in the closer fit to the true composition distribution, but still suffers considerable
dispersion in the probabilities (in particular, the overlap in fitted probabilities
between the composition groups is large relative to the mean displacement between groups).

The results of the Mann-Whitney U test show the impact, collectively, of changing
composition and morphology of the high-$Y_e$ ejecta.
However, these group partitions into statistically significant differences in the
distribution of observables do not provide a sufficient subset of observables
to categorize new kilonova data points using the IRLS method for logistic regression.
Since we have restricted our consideration to two angular views and three times of the
$g$ band, $K$ band and bolometric luminosity, we may have excluded other observables
that are significant indicators of morphology and composition.
We have tested several different sub-vectors per categorization test, but
note that the possibilities of sub-vector combinations can be expanded by
using more viewing angles, broadband magnitudes, or times.
Another factor that complicates the categorization of these models by composition or
morphology is the diversity of the light curves over the other properties: viewing angle,
high-$Y_e$ ejecta mass and velocity, and low-$Y_e$ ejecta mass and velocity.
We are left to conclude that either a different fitting method over a different sub-vector
of data is required, or that categorization by morphology or composition is inefficient
due to the strong effects of the other model properties.

\subsection{Comparison with AT2017gfo}
\label{sec:at17}

We next compare AT2017gfo to our model data set using a logistic regression, as
described in Section~\ref{sec:irls}.
For the sub-vector of data, we use the axial and edge luminosities at days
1, 4, and 8, given their low $p$-value from the Mann-Whitney U test in
Section~\ref{sec:mwut}.
Performing a logistic regression to fit parameter coefficients using just bolometric
luminosity from the model data set, and evaluating the resulting function at the
observed bolometric luminosity for AT2017gfo from~\cite{smartt2017}, we find a
probability estimate of $\sim$65\% for a high-$Y_e$ component with a wind 2
composition, whereas the mean fitted probability of the data set from this
regression is $\sim$46\% for wind 1 models being categorized as wind 2
and $\sim$54\% for wind 2 models being categorized (correctly) as wind 2.
The goodness of fit is low, given the close mean probabilities, but the result
suggests more closely examining models with high-$Y_e$ components that have wind 2
compositions.

If the probability from the Mann-Whitney U test-based IRLS logistic regression
is taken to reduce the consideration to wind 2 compositions, that still leaves
450 models to consider.
It may be worth reducing the number of models considered further with another
metric that is simpler than the full spectra at each observed time; for instance
one possible metric is an L$_1$ error measure of the bolometric luminosity,
\begin{equation}
  \label{eq1:at17}
  \varepsilon_{\theta} =
  \frac{1}{N}\sum_i^N\frac{|L_{\theta}(t_i)-L_{\rm obs,i}|}{L_{\rm obs,i}} \;\;,
\end{equation}
where $\varepsilon_{\theta}$ is the model error at angle $\theta$ off axis,
$N$ is the number of observed time points, $L_{\theta}$ is the model bolometric
luminosity at $\theta$, and $L_{\rm obs,i}$ is the observed luminosity at time $t_i$.
If we have an independent constraint that the observer angle is $\sim18^o$ off-axis,
we may restrict the angular views considered for Eq.~\eqref{eq1:at17}.
This particular choice of observer angle is motivated by recent estimation for
GW170817, ${\theta\approx15^\circ-22^\circ}$~\citep{hotokezaka2019}.
Table~\ref{tb1:at17} has the partitioning of the number of models by the error
calculated with Eq.~\eqref{eq1:at17} in the angular viewing bin containing $18^o$.

\begin{table}
  \centering
  \caption{L$_1$ error ranges in bolometric luminosity and corresponding number
    of models with wind 2 composition for the high-$Y_e$ component.}
  \begin{tabular}{cc}
    \hline
    $\epsilon_{M,18^o}$ & Number of Wind 2 Models \\
    \hline
    $\geq$ 1.0 & 329 \\
    $\geq$ 0.9, $<$ 1.0 & 24 \\
    $\geq$ 0.8, $<$ 0.9 & 21 \\
    $\geq$ 0.7, $<$ 0.8 & 16 \\
    $\geq$ 0.6, $<$ 0.7 & 16 \\
    $\geq$ 0.5, $<$ 0.6 & 16 \\
    $\geq$ 0.4, $<$ 0.5 & 20 \\
    $\geq$ 0.3, $<$ 0.4 & 8 \\
    $\geq$ 0.2, $<$ 0.3 & 0 \\
    $<$ 0.2 & 0 \\
    \hline
  \end{tabular}
  \label{tb1:at17}
\end{table}

We may then examine models from a few different luminosity error groups, for
instance the models and corresponding luminosity errors shown in Table~\ref{tb2:at17}.
We have introduced an off-grid model with an average high-$Y_e$ ejecta velocity of
0.2$c$ for an additional comparison.
The error in overall bolometric luminosity appears to systematically decrease as
the ejecta velocity is increased, which would suggest that higher velocity is favored
for AT2017gfo at the given viewing angle of $\sim18^o$.
However, in Table~\ref{tb2:at17} we have constrained the other ejecta
properties to specific values, so this error trend does not generally hold for
all models.

\begin{table}
  \centering
  \caption{Sample models and corresponding bolometric luminosity error.}
  \begin{tabular}{cc}
    \hline
    Model & $\epsilon_{M,18^o}$ \\
    \hline
    T\_m0.01v0.3\_P2\_m0.03v0.05 & 1.24 \\
    T\_m0.01v0.3\_P2\_m0.03v0.15 & 0.77 \\
    T\_m0.01v0.3\_P2\_m0.03v0.2  & 0.70 \\
    T\_m0.01v0.3\_P2\_m0.03v0.3  & 0.45 \\
    \hline
  \end{tabular}
  \label{tb2:at17}
\end{table}

\begin{figure*}
  \centering
  \includegraphics[width=0.9\columnwidth]{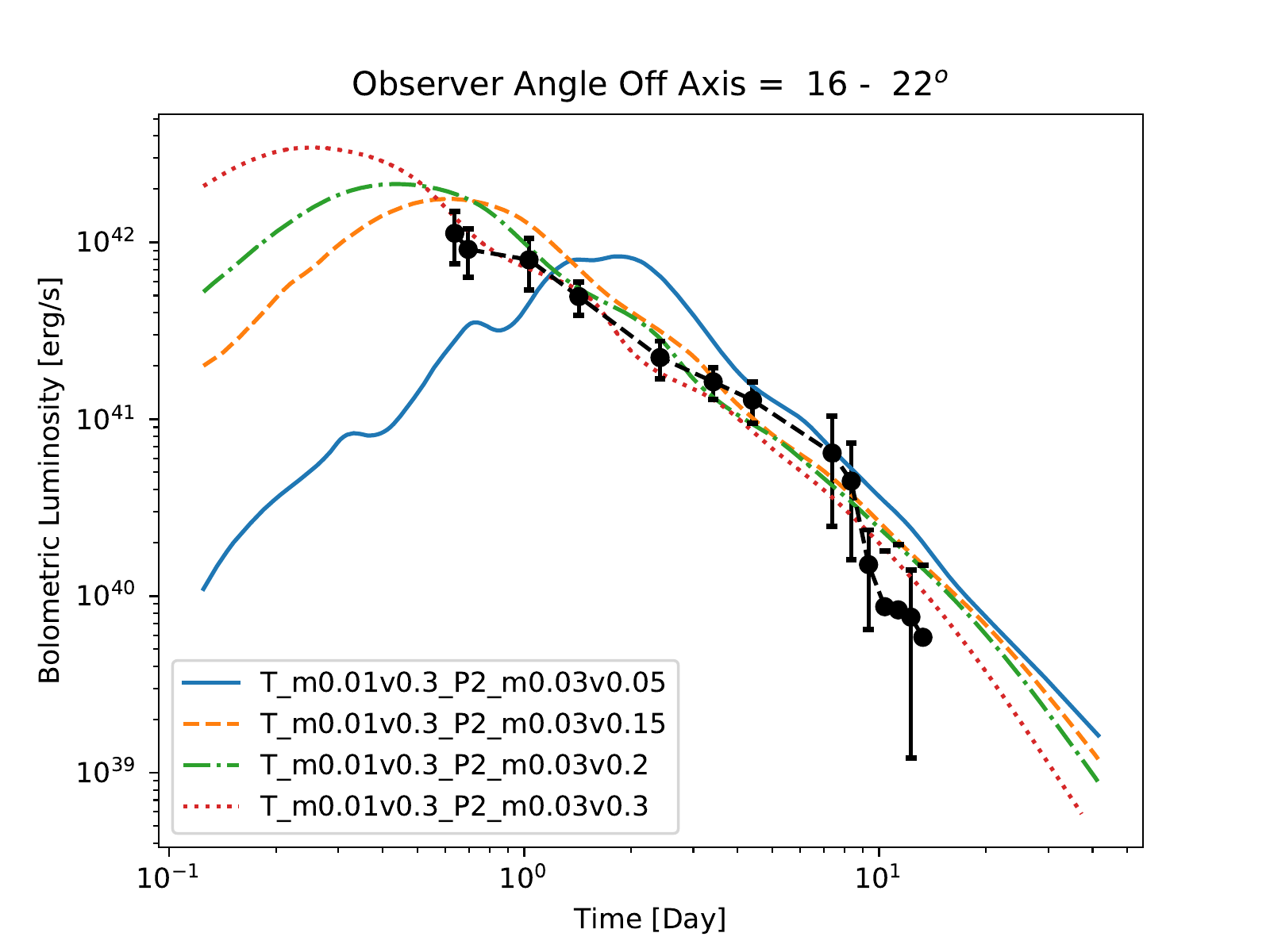}
  \includegraphics[width=0.9\columnwidth]{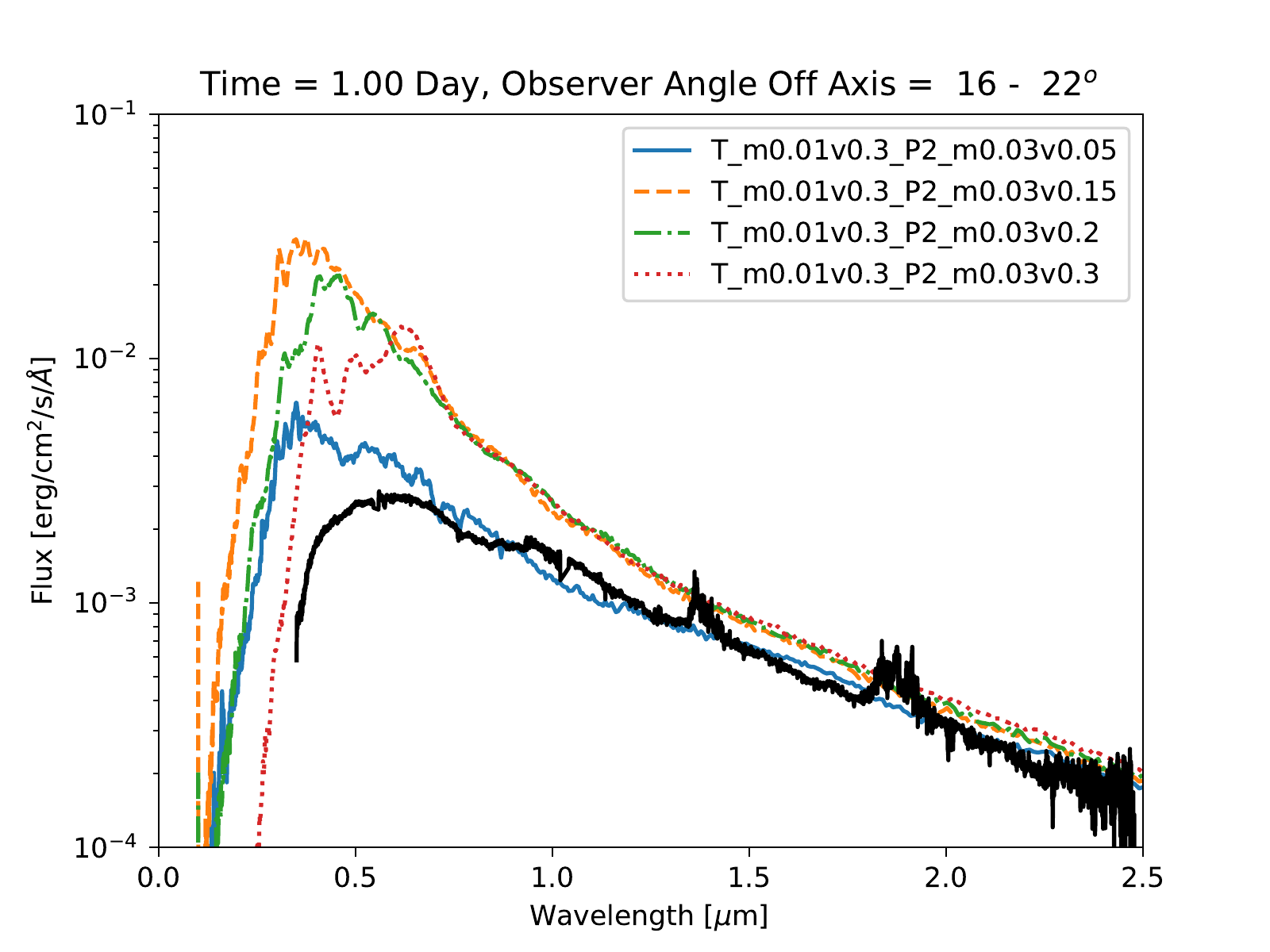} \\
  \includegraphics[width=0.9\columnwidth]{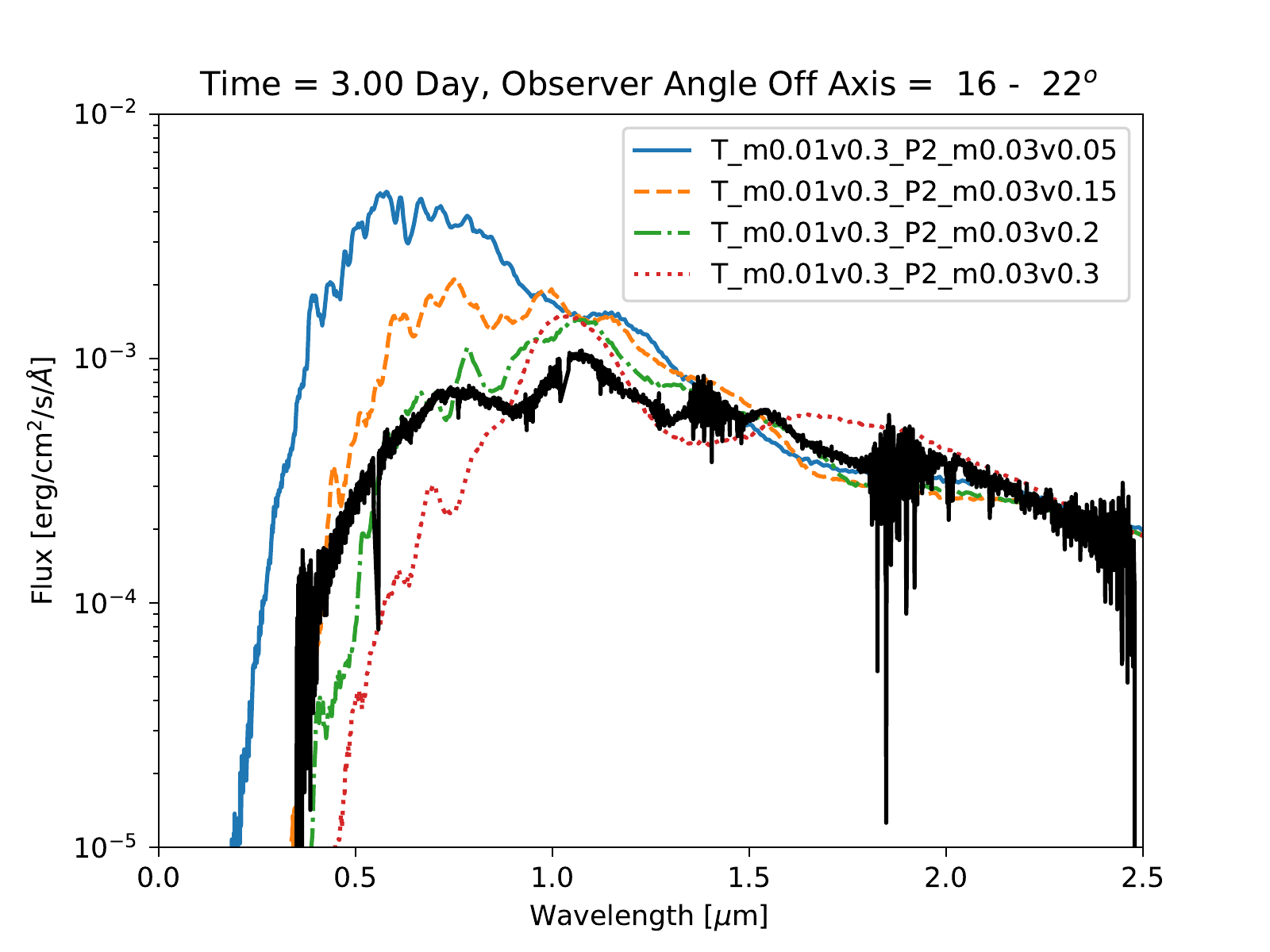}
  \includegraphics[width=0.9\columnwidth]{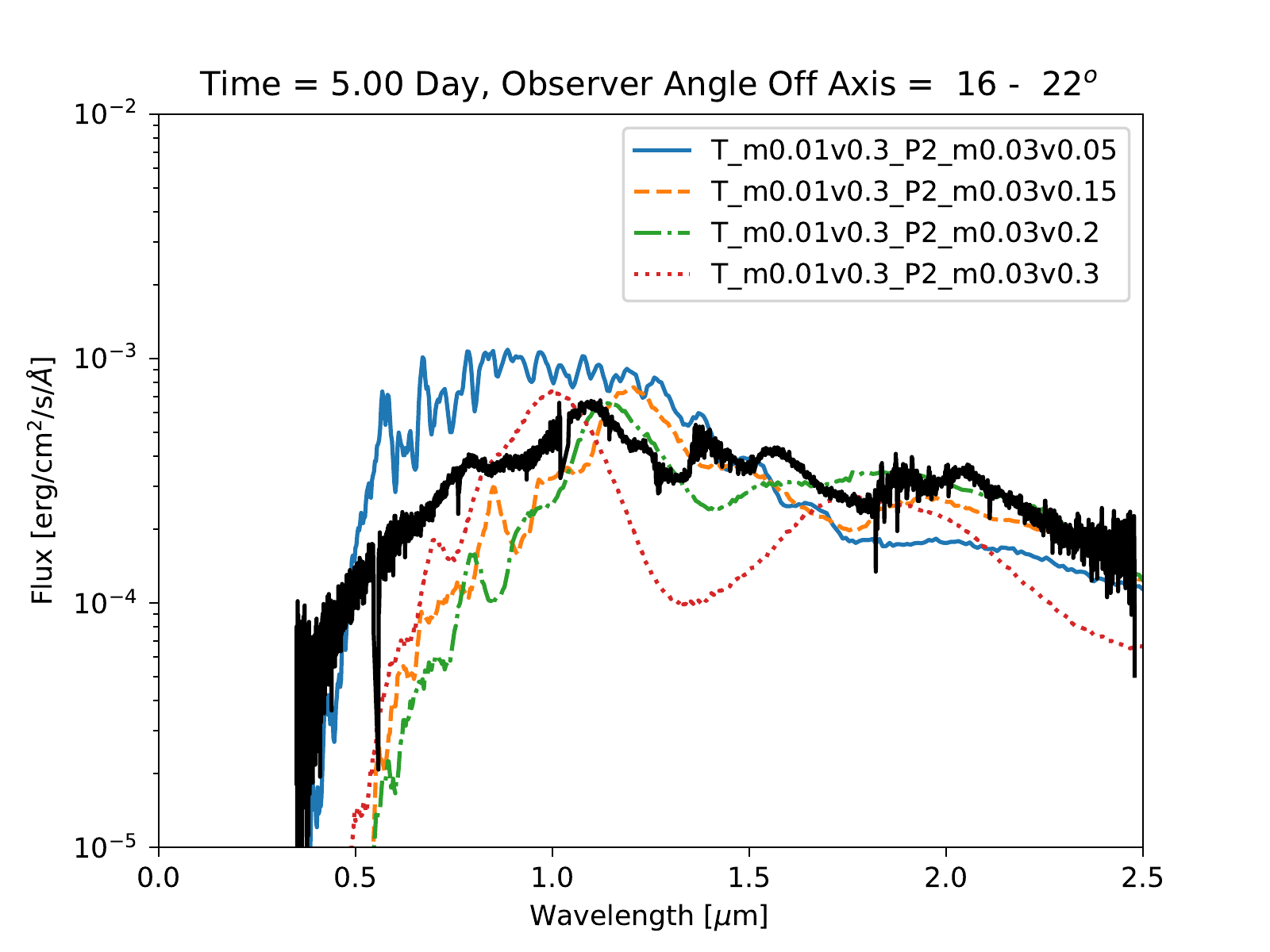}
  \caption{Luminosity versus time (top left), spectra versus wavelength
    at day 1 (top right), day 3 (bottom left), and day 5 (bottom right).
    All data are shown at an estimated observer angle of
    ${\theta\approx15^\circ-22^\circ}$ \citep{hotokezaka2019}.}
  \label{fg1:at17}
\end{figure*}

Figure~\ref{fg1:at17} has bolometric luminosity versus time for the models
listed in Table~\ref{tb2:at17} along with the AT2017gfo data from~\cite{smartt2017}
(top left panel), and spectra versus wavelength at days 1 (top right), 3 (bottom left)
and 5 (bottom right) along with VLT/X-shooter data presented by~\cite{pian2017}.
Despite having the lowest error in Table~\ref{tb2:at17}, the model with an average
high-$Y_e$ ejecta velocity of 0.3$c$ produces a broad absorption feature between
1 and 2~$\mu$m at day 5 that is not present in the observed spectrum.

This behavior implies that a more complicated error metric may be needed to filter data, if
such a step is taken before comparing spectra.
This may also imply that there is some tension between the observation and the model,
and that a fit should account for both consistency in spectra and in bolometric luminosity.
Notably, the models with high-$Y_e$ ejecta velocity $>0.05c$ are all dim in the
optical range relative to AT2017gfo by day 5; this is a consistent feature of
our models, which warrants further investigation into both the numerical method
and model assumptions.

\section{Conclusions}

We have simulated a broad grid of 900 2-component 2D axisymmetric kilonova models
in order to supply a basis for model analysis and comparison to observations.
Each ejecta component ranges from 0.001 to 0.1 M$_{\odot}$ in mass and 0.05 to
0.3$c$ in average ejecta speed.
The model grid can support constraining the range of ejected
mass from an observation or upper bound on an observation; we have indeed applied
these models for the observational studies of~\cite{thakur2020},~\cite{oconnor2021}
and~\cite{bruni2021}.
We emphasize as a caveat that the grid does not exhaust the space of possible
kilonova models.
For instance, this grid is not as varied in morphology as that of~\cite{korobkin2020}
or as varied in composition as that of~\cite{even2019}.
Examples from these studies that we have not included in this grid are
non-toroidal low-$Y_e$ ejecta and solar abundance r-process composition.
Moreover, as a consequence of uncertainties in the nuclear mass models, there
are uncertainties in the heating rate due to r-process
\citep{barnes2020,zhu2021} that we have not explored in this model grid.
There has also been recent progress in non-LTE physics~\citep{hotokezaka2021};
we have not included any non-LTE treatment in the models presented here.
Thus comparisons of observations to the model grid are limited by these model
constraints, mass model uncertainties, and limitations in fidelity.

We have explored some example uses of the model grid in this work, including
basic population statistics and comparison to the spectra of AT2017gfo.
Below is a summary of this work and the corresponding results.

\begin{itemize}
\item The models have been simulated with multifrequency LTE radiative transfer
  and detailed opacities, along with detailed radioactive heating based on
  nucleosynthetic tracer points~\citep{wollaeger2018} and the thermalization
  fraction formulation of~\cite{barnes2016}.
  These models cover a large range of masses and velocities from the literature,
  but are restricted to toroidal low-$Y_e$ ejecta and either spherical or lobed
  high-$Y_e$ ejecta morphologies~\citep{korobkin2020}.
\item Axial and edge view $g$ band, $K$ band, and bolometric luminosity values
  at days 1, 4 and 8 are listed in the expanded form of Table~\ref{tb1:app}.
  We have showcased these data in an example involving the non-parametric
  Mann-Whitney U test for determining whether different partitions (or grouping) by
  properties produce statistically distinct distributions of observable properties.
  For this model set, the results imply that grouping by high-$Y_e$ composition is more
  effective in producing distinct distributions than grouping by high-$Y_e$ morphology.
  We emphasize that this test is possible for this model grid because there are
  only two morphologies and two compositions for the high-$Y_e$ component.
  A model set more exhaustive in morphology and composition would complicate this
  test, and generally minimize its effectiveness.
\item Using observables over which the Mann-Whitney U test produces statistically
  distinct distributions in a logistic regression produces a low-quality fit,
  indicating more data (than presented in expanded Table~\ref{tb1:app}) or
  more sophisticated techniques are needed to properly categorize observations with
  respect to the model data.
  We do find that the logistic regression performs better for the grouping by
  composition than for the grouping by morphology, which is consistent with the
  Mann-Whitney U tests.
\item We have compared some of our model light curves and spectra from this grid
  to AT2017gfo, and find that an overly simple bolometric luminosity error estimate
  may filter out models that have better spectral agreement than lower-error models.
  The model spectra shown for high-$Y_e$ velocity $>0.05c$ exhibit a spectral cliff at
  late time, as discussed by~\cite{wollaeger2018}, that is not present in the late
  spectrum of AT2017gfo.
\end{itemize}

The statistical study presented here can be further expanded using machine
learning (ML) classification methods (see, for instance,~\citet{bishop1995,zhang2000}).
We consider more advanced ML techniques for light curve interpolation in another
work~\citep{ristic2021}.
Interpolation should permit sparser initial grid requirements, which can also be
achieved with Latin Hypercube sampling (see, for example,~\citet{stein1987}).
The grid we have presented here is uniform, but could be extended via
Latin Hypercube sampling to account for new parameter variation.
One potential impediment to Latin Hypercube sampling is the lack of an
informed joint probability distribution over the model input.
More observations, ab initio merger simulations, and population synthesis may
help to estimate this probability distribution and enable efficient Latin
Hypercube sampling.

\section*{Acknowledgements}

This work was supported by the US Department of Energy through the Los Alamos National Laboratory.
Los Alamos National Laboratory is operated by Triad National Security, LLC, for the
National Nuclear Security Administration of U.S.\ Department of Energy (Contract No.\ 89233218CNA000001).
Research presented in this article was supported by the Laboratory Directed Research and Development
program of Los Alamos National Laboratory under project number 20190021DR.
This research used resources provided by the Los Alamos National Laboratory Institutional Computing
Program, which is supported by the U.S. Department of Energy National Nuclear Security Administration
under Contract No.\ 89233218CNA000001.
EAC acknowledges financial support from the IDEAS Fellowship, a research traineeship program funded by the National Science Foundation under grant DGE-1450006.
ROS and MR acknowledge support from NSF AST 1909534.

\bibliography{refs}

\appendix

\section{Magnitude Tables}
\label{sec:app}

\begin{table*}
  \centering
  \scriptsize
  \caption{Axial / edge viewing bin for absolute AB magnitudes in g, K bands and
    bolometric luminosity ($L_{\rm bol}$ [$10^{40}$ erg/s]) for models with 0.001 M$_{\odot}$
    dynamica ejecta and P morphology, high-latitude composition (1) wind.}
  \begin{tabular}{l|ccc|ccc|ccc}
  \hline\hline
  & \multicolumn{3}{|c}{Day 1} & \multicolumn{3}{|c}{Day 4} & \multicolumn{3}{|c}{Day 8} \\
  Model & g & K & $L_{\rm bol}$ & g & K & $L_{\rm bol}$ & g & K & $L_{\rm bol}$ \\
  \hline
  T\_m0.001v0.05\_P1\_m0.001v0.05         & -10.8 / -11.1         & -12.2 / -11.5         & 2.0 / 1.6             & -0.7 / -1.0           & -12.1 / -11.2         & 0.4 / 0.2             & 4.0 / 4.3             & -10.4 / -9.8              & 0.1 / 0.1  \\
  T\_m0.001v0.05\_P1\_m0.001v0.15         & -7.5 / -9.8           & -12.2 / -11.3         & 1.1 / 1.1             & 0.1 / 1.0             & -11.8 / -10.8         & 0.3 / 0.2             & 1.5 / 4.3             & -10.3 / -9.6              & 0.1 / 0.1  \\
  T\_m0.001v0.05\_P1\_m0.001v0.3          & -5.9 / -5.5           & -12.4 / -11.8         & 1.1 / 0.9             & -2.5 / 2.4            & -11.5 / -10.6         & 0.2 / 0.1             & -0.8 / 4.6            & -10.0 / -9.5              & 0.1 / 0.0  \\
  T\_m0.001v0.05\_P1\_m0.003v0.05         & -12.9 / -13.3         & -12.2 / -11.6         & 5.1 / 5.5             & -1.8 / -2.4           & -12.2 / -11.6         & 0.5 / 0.3             & 3.2 / 2.7             & -10.6 / -10.1             & 0.1 / 0.1  \\
  T\_m0.001v0.05\_P1\_m0.003v0.15         & -10.6 / -11.1         & -12.2 / -11.7         & 2.1 / 3.1             & -0.9 / 0.2            & -11.9 / -11.0         & 0.4 / 0.2             & 0.6 / 3.5             & -10.3 / -9.7              & 0.1 / 0.1  \\
  T\_m0.001v0.05\_P1\_m0.003v0.3          & -7.4 / -7.0           & -12.5 / -11.7         & 1.5 / 1.3             & -3.8 / 1.3            & -11.6 / -10.6         & 0.3 / 0.1             & -2.2 / 3.6            & -10.0 / -9.5              & 0.1 / 0.1  \\
  T\_m0.001v0.05\_P1\_m0.01v0.05          & -13.9 / -14.2         & -12.4 / -12.2         & 10.7 / 12.4           & -5.0 / -5.5           & -12.5 / -11.5         & 0.8 / 0.6             & 2.0 / 2.2             & -11.2 / -11.0             & 0.2 / 0.1  \\
  T\_m0.001v0.05\_P1\_m0.01v0.15          & -12.8 / -13.1         & -12.3 / -12.4         & 6.4 / 9.8             & -2.0 / -0.8           & -12.1 / -12.0         & 0.5 / 0.5             & -0.4 / 2.5            & -10.4 / -10.0             & 0.1 / 0.1  \\
  T\_m0.001v0.05\_P1\_m0.01v0.3           & -10.8 / -10.7         & -12.5 / -11.7         & 3.6 / 6.4             & -5.0 / 0.3            & -11.6 / -10.7         & 0.4 / 0.3             & -3.7 / 2.3            & -10.0 / -9.5              & 0.1 / 0.1  \\
  T\_m0.001v0.05\_P1\_m0.03v0.05          & -14.2 / -14.5         & -12.6 / -12.6         & 15.0 / 17.6           & -8.8 / -9.3           & -13.3 / -12.6         & 2.9 / 2.9             & 0.8 / 0.3             & -12.0 / -12.2             & 0.4 / 0.4  \\
  T\_m0.001v0.05\_P1\_m0.03v0.15          & -14.7 / -15.0         & -12.0 / -12.5         & 19.0 / 29.5           & -3.3 / -2.7           & -12.9 / -13.3         & 1.1 / 1.4             & -1.3 / 1.4            & -11.2 / -11.0             & 0.3 / 0.3  \\
  T\_m0.001v0.05\_P1\_m0.03v0.3           & -14.2 / -13.3         & -12.7 / -13.1         & 18.8 / 22.1           & -6.0 / -0.5           & -12.1 / -11.4         & 0.9 / 0.7             & -4.9 / 1.1            & -10.1 / -9.6              & 0.2 / 0.2  \\
  T\_m0.001v0.05\_P1\_m0.1v0.05           & -14.3 / -14.6         & -12.9 / -12.9         & 17.1 / 20.3           & -12.0 / -12.2         & -14.0 / -14.0         & 10.4 / 12.8           & -3.2 / -3.8           & -13.1 / -13.2             & 1.3 / 1.5  \\
  T\_m0.001v0.05\_P1\_m0.1v0.15           & -15.9 / -16.3         & -12.6 / -13.6         & 55.7 / 80.7           & -7.1 / -7.9           & -13.7 / -12.8         & 3.3 / 4.8             & -2.3 / 0.4            & -12.7 / -12.9             & 0.8 / 0.9  \\
  T\_m0.001v0.05\_P1\_m0.1v0.3            & -15.7 / -15.0         & -13.6 / -13.9         & 65.3 / 59.5           & -7.1 / -2.1           & -13.4 / -13.9         & 2.8 / 3.5             & -6.0 / 0.3            & -10.9 / -10.6             & 0.6 / 0.5  \\
  T\_m0.001v0.15\_P1\_m0.001v0.05         & -11.0 / -6.7          & -13.2 / -12.3         & 3.4 / 0.8             & 2.2 / 0.3             & -11.3 / -10.7         & 0.2 / 0.1             & 4.9 / 0.0             & -9.8 / -9.4               & 0.0 / 0.0  \\
  T\_m0.001v0.15\_P1\_m0.001v0.15         & -8.4 / -8.7           & -13.2 / -12.3         & 2.4 / 1.4             & 0.2 / 2.7             & -11.2 / -10.6         & 0.2 / 0.1             & 1.5 / 5.3             & -9.8 / -9.4               & 0.0 / 0.0  \\
  T\_m0.001v0.15\_P1\_m0.001v0.3          & -5.9 / -6.5           & -13.3 / -12.5         & 1.8 / 1.2             & -2.5 / 2.9            & -11.0 / -10.4         & 0.1 / 0.1             & -0.8 / 4.7            & -9.4 / -9.2               & 0.0 / 0.0  \\
  T\_m0.001v0.15\_P1\_m0.003v0.05         & -13.4 / -8.1          & -13.3 / -12.3         & 9.0 / 0.9             & 0.2 / 1.2             & -11.7 / -11.3         & 0.3 / 0.2             & 4.3 / 0.6             & -10.1 / -9.8              & 0.1 / 0.1  \\
  T\_m0.001v0.15\_P1\_m0.003v0.15         & -10.9 / -10.8         & -13.4 / -12.5         & 4.2 / 3.2             & -0.7 / 1.6            & -11.4 / -10.8         & 0.2 / 0.2             & 0.6 / 3.6             & -9.9 / -9.5               & 0.1 / 0.1  \\
  T\_m0.001v0.15\_P1\_m0.003v0.3          & -7.4 / -7.4           & -13.4 / -12.5         & 2.3 / 1.7             & -3.8 / 1.9            & -11.0 / -10.4         & 0.2 / 0.1             & -2.3 / 3.8            & -9.5 / -9.2               & 0.0 / 0.0  \\
  T\_m0.001v0.15\_P1\_m0.01v0.05          & -14.5 / -8.8          & -13.4 / -12.4         & 20.1 / 1.0            & -4.3 / 0.9            & -12.5 / -11.9         & 0.7 / 0.3             & 1.8 / 0.6             & -10.9 / -10.8             & 0.1 / 0.1  \\
  T\_m0.001v0.15\_P1\_m0.01v0.15          & -13.0 / -13.0         & -13.6 / -12.9         & 9.9 / 9.1             & -1.9 / 0.3            & -12.1 / -11.9         & 0.5 / 0.4             & -0.3 / 2.8            & -10.2 / -10.0             & 0.1 / 0.1  \\
  T\_m0.001v0.15\_P1\_m0.01v0.3           & -10.8 / -10.7         & -13.6 / -12.7         & 4.8 / 6.9             & -5.0 / 0.7            & -11.2 / -10.7         & 0.3 / 0.3             & -3.7 / 2.5            & -9.6 / -9.3               & 0.1 / 0.1  \\
  T\_m0.001v0.15\_P1\_m0.03v0.05          & -14.8 / -9.2          & -13.6 / -12.5         & 26.5 / 1.2            & -9.0 / -4.0           & -13.4 / -12.7         & 2.9 / 0.6             & 0.9 / 0.7             & -11.9 / -12.1             & 0.3 / 0.3  \\
  T\_m0.001v0.15\_P1\_m0.03v0.15          & -14.9 / -14.9         & -13.7 / -13.1         & 26.4 / 26.7           & -3.3 / -2.3           & -13.3 / -13.4         & 1.4 / 1.4             & -1.2 / 2.1            & -11.2 / -11.1             & 0.3 / 0.2  \\
  T\_m0.001v0.15\_P1\_m0.03v0.3           & -14.2 / -13.3         & -13.8 / -13.6         & 20.2 / 22.4           & -6.0 / -0.2           & -12.0 / -11.6         & 0.8 / 0.7             & -4.9 / 1.6            & -10.0 / -9.7              & 0.2 / 0.2  \\
  T\_m0.001v0.15\_P1\_m0.1v0.05           & -14.8 / -9.4          & -13.7 / -12.6         & 27.9 / 1.3            & -12.6 / -8.7          & -14.5 / -14.0         & 15.5 / 3.1            & -3.7 / 0.6            & -13.3 / -13.5             & 1.5 / 1.2  \\
  T\_m0.001v0.15\_P1\_m0.1v0.15           & -16.1 / -16.1         & -13.9 / -13.8         & 69.6 / 68.4           & -8.0 / -8.1           & -14.2 / -13.3         & 5.4 / 4.9             & -2.2 / 1.1            & -12.9 / -13.0             & 0.9 / 0.9  \\
  T\_m0.001v0.15\_P1\_m0.1v0.3            & -15.7 / -14.9         & -14.3 / -14.3         & 66.9 / 57.6           & -7.1 / -1.9           & -13.5 / -14.0         & 2.7 / 3.6             & -6.0 / 0.5            & -11.1 / -10.8             & 0.6 / 0.5  \\
  T\_m0.001v0.3\_P1\_m0.001v0.05          & -9.8 / -1.1           & -13.4 / -12.6         & 2.2 / 0.7             & 1.7 / 0.3             & -10.7 / -10.1         & 0.1 / 0.1             & 3.8 / 0.0             & -8.4 / -7.8               & 0.0 / 0.0  \\
  T\_m0.001v0.3\_P1\_m0.001v0.15          & -6.9 / -5.1           & -13.3 / -12.6         & 1.5 / 0.7             & 0.4 / 4.3             & -10.5 / -9.8          & 0.1 / 0.1             & 1.6 / 5.6             & -8.3 / -7.6               & 0.0 / 0.0  \\
  T\_m0.001v0.3\_P1\_m0.001v0.3           & -5.9 / -2.6           & -13.2 / -12.6         & 1.2 / 0.8             & -2.5 / 3.5            & -10.5 / -9.7          & 0.1 / 0.1             & -0.8 / 5.3            & -8.1 / -7.5               & 0.0 / 0.0  \\
  T\_m0.001v0.3\_P1\_m0.003v0.05          & -12.8 / -1.7          & -13.6 / -12.6         & 7.2 / 0.7             & 0.9 / 0.1             & -11.2 / -10.9         & 0.1 / 0.1             & 5.0 / 0.0             & -8.9 / -8.6               & 0.0 / 0.0  \\
  T\_m0.001v0.3\_P1\_m0.003v0.15          & -10.6 / -8.1          & -13.5 / -12.7         & 3.9 / 1.1             & -0.6 / 2.3            & -10.7 / -10.1         & 0.1 / 0.1             & 0.8 / 5.1             & -8.5 / -7.8               & 0.0 / 0.0  \\
  T\_m0.001v0.3\_P1\_m0.003v0.3           & -7.5 / -6.2           & -13.4 / -12.6         & 2.1 / 1.2             & -3.8 / 2.5            & -10.5 / -9.8          & 0.1 / 0.1             & -2.2 / 4.2            & -8.2 / -7.6               & 0.0 / 0.0  \\
  T\_m0.001v0.3\_P1\_m0.01v0.05           & -14.4 / -2.9          & -13.8 / -12.7         & 18.8 / 0.8            & -2.6 / 0.4            & -12.2 / -12.2         & 0.4 / 0.4             & 2.1 / 0.0             & -10.2 / -10.2             & 0.1 / 0.1  \\
  T\_m0.001v0.3\_P1\_m0.01v0.15           & -13.0 / -10.8         & -14.0 / -13.1         & 12.1 / 3.0            & -1.8 / 2.2            & -11.6 / -11.4         & 0.3 / 0.3             & 0.0 / 4.5             & -9.0 / -8.7               & 0.1 / 0.1  \\
  T\_m0.001v0.3\_P1\_m0.01v0.3            & -10.9 / -10.4         & -13.7 / -12.8         & 5.9 / 5.7             & -5.0 / 1.5            & -10.9 / -10.3         & 0.2 / 0.2             & -3.7 / 3.1            & -8.5 / -7.8               & 0.0 / 0.0  \\
  T\_m0.001v0.3\_P1\_m0.03v0.05           & -14.7 / -3.1          & -13.9 / -12.8         & 25.6 / 0.9            & -6.7 / 0.9            & -13.5 / -13.0         & 2.1 / 1.0             & -0.4 / 0.0            & -11.5 / -11.8             & 0.2 / 0.2  \\
  T\_m0.001v0.3\_P1\_m0.03v0.15           & -14.9 / -13.3         & -14.4 / -13.3         & 34.7 / 9.0            & -2.9 / 0.9            & -13.2 / -13.3         & 1.1 / 1.1             & -0.9 / 2.4            & -10.4 / -10.4             & 0.2 / 0.2  \\
  T\_m0.001v0.3\_P1\_m0.03v0.3            & -14.2 / -13.0         & -14.2 / -13.8         & 23.9 / 19.6           & -6.0 / 0.5            & -12.1 / -11.5         & 0.8 / 0.6             & -4.9 / 2.1            & -9.1 / -8.5               & 0.2 / 0.1  \\
  T\_m0.001v0.3\_P1\_m0.1v0.05            & -14.1 / -2.1          & -14.0 / -12.9         & 19.1 / 1.0            & -11.1 / -3.6          & -14.4 / -14.3         & 10.7 / 3.6            & -2.7 / 0.0            & -12.9 / -13.3             & 0.9 / 1.2  \\
  T\_m0.001v0.3\_P1\_m0.1v0.15            & -16.3 / -14.7         & -14.6 / -13.9         & 93.4 / 23.3           & -6.8 / -4.8           & -14.6 / -14.1         & 6.0 / 3.7             & -2.0 / 1.3            & -12.5 / -12.5             & 0.7 / 0.7  \\
  T\_m0.001v0.3\_P1\_m0.1v0.3             & -15.7 / -14.7         & -14.8 / -14.6         & 76.3 / 51.3           & -7.1 / -1.7           & -13.9 / -14.1         & 3.3 / 3.8             & -6.0 / 0.9            & -10.8 / -10.4             & 0.6 / 0.5  \\
  \hline
\end{tabular}
\label{tb1:app}
\end{table*}

\end{document}